\documentclass[12pt]{JHEP3}
\usepackage{amsmath,amssymb,graphics,amscd,amsfonts,mathrsfs,epsf,indentfirst,cite}

\allowdisplaybreaks[1]
\usepackage{epsfig}
\usepackage{bm}
\usepackage{subfigure}

\title{
Conic D-branes
}

\author{Koji Hashimoto$^{1}$, Shunichiro Kinoshita$^{2}$ and Keiju Murata$^{3}$
\\
$^{1}${\it Department of Physics, Osaka University,
Toyonaka, Osaka 560-0043, Japan}
\\
$^{2}${\it Department of Physics, Chuo University, Tokyo 112-8551, Japan}
\\
$^{3}${\it Keio University, 4-1-1 Hiyoshi, Yokohama 223-8521, Japan}
\\
\\
{\it E-mails:} \email{koji@phys.sci.osaka-u.ac.jp},
\email{kinoshita@phys.chuo-u.ac.jp},
\email{keiju@phys-h.keio.ac.jp}
}

\abstract{
The shape of D-branes is of fundamental interest in string theory.
We find that generically D-branes in trivial spacetime can form a conic shape 
under external uniform forces. Surprisingly, the apex angle is found to be unique, once
the spatial dimensions of the cone is given. In particular
it is universal irrespective of the external forces. The quantized angle is reminiscent 
of Taylor cones of hydrodynamic electrospray.
We provide explicit D-brane solutions as well as 
the mechanism of a force balance on the cone, for D-branes 
in RR and NSNS flux backgrounds. Critical embedding of probe D-branes 
in AdS/CFT with electric and magnetic fields is in the same category, for which
we give an analytic proof of 
a power-low spectrum of ``turbulent meson condensation.''
}

\preprint{
{\normalsize OU-HET-861} 
}


\begin{document}

\maketitle

\section{Introduction}
\label{sec:intro}

Spiky branes are of fundamental interest in string theory and M-theory, since the issue of
membrane quantization has an obstacle of the spike singularity \cite{deWit:1988ct}. 
For membranes there exists deformations of their surfaces to have thin spike without costing increase of the volume, thus suffers from infinite degeneracies resulting in a continuous
spectrum. A resolution of this fundamental issue would need some findings about how
spiky/conic branes can be stabilized classically. 

Based on this motivation, in this paper we find new conic D-brane configurations in the background flux. They are solutions of classical D-brane effective actions, which are Dirac-Born-Infeld (DBI) actions. Surprisingly, the apex angle of the cone is found to be universal
and depends only on the dimensions of the cone worldvolume. 

Note that the D-brane cones we found do not use nontrivial topologies of background spacetime in which the D-branes are embedded. D-branes wrapping conifolds or orbifolds
have been studied in various context in string theory, while ours are not of that kind.
In the target space, the tip of our D-brane cones is not located at special points (such as the tips of the conifolds or black hole horizons). Our examples include a 
previously known probe D-brane configuration in AdS/CFT correspondence
where a critical value of an electric field on the brane forces  it to have a conic shape,
which is called critical embedding \cite{Erdmenger:2007bn,Albash:2007bq}\footnote{
This critical embedding with the electric field in AdS/CFT is different from
the critical embedding at thermal phase transition \cite{Mateos:2006nu,Frolov:2006tc} 
where the apex of the probe D-brane touches event horizon of a background 
black hole spacetime. As we emphasized, the conic D-branes of our interest
are formed without the help of nontrivial background spacetime.}.

We are motivated by a hydrodynamic phenomena called Taylor cones \cite{Taylor} which are widely used in electrospray in material/industrial sciences (see Fig.~\ref{Fig:Taylor}). 
The Taylor cones are generically 
formed for charged surfaces of liquid under some background electric field. The mechanism is quite simple: the induced charges on the surface of the liquid repel each other and cancel
the surface tension, and the instability grows to form a cone whose apex has a vanishing tension due to the cancellation. Interestingly, the Taylor cones
are popular for their universal cone angle $\theta_{\rm cone}=49.29^{\rm o}$, which
can be easily derived from the
tension valance on the surface and with the Maxwell equations.
The simple reasoning of the Taylor cones lead us to the new examples of D-brane cones
presented in this paper.

%
\FIGURE{ 
\includegraphics[width=13cm]{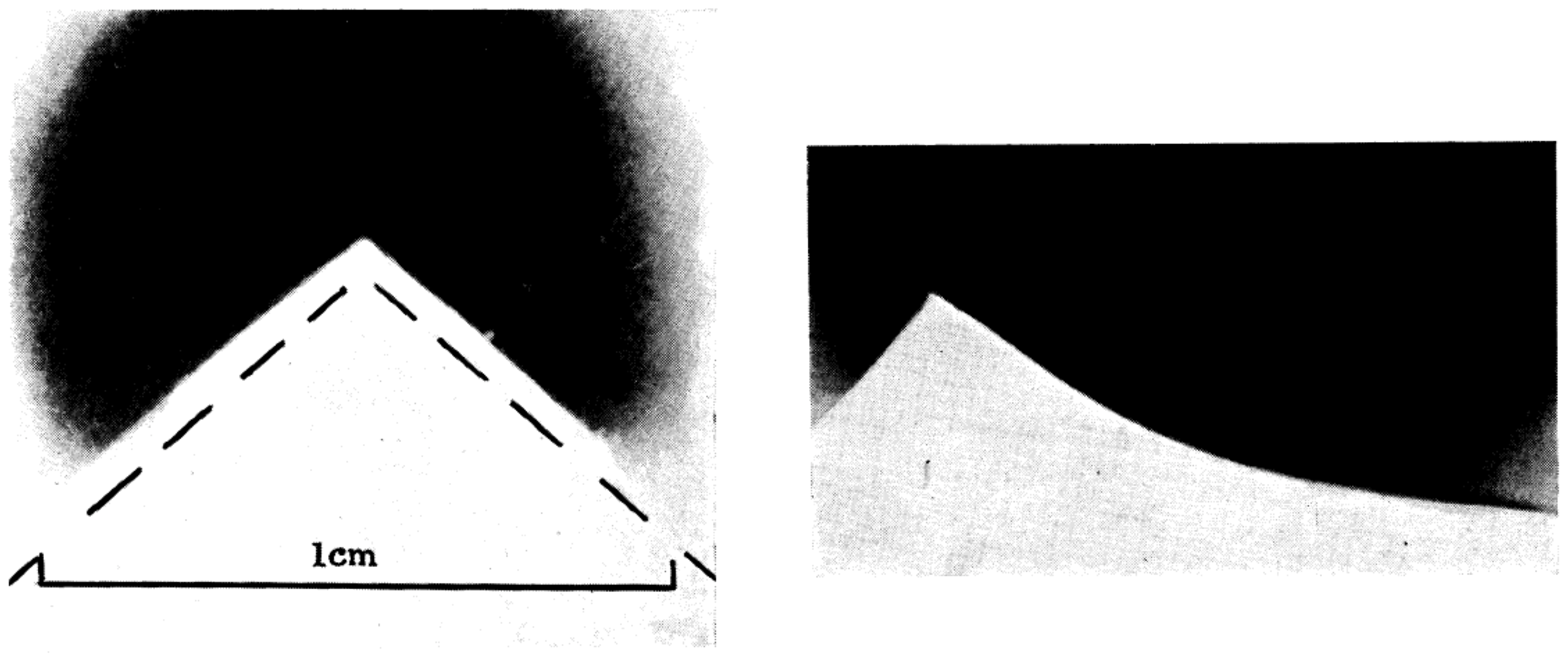}
\caption{Photos of Taylor cones formed at oil/water interface in 
experiments \cite{Taylor}. Left: Pointed summit formed under an electric field along 
the vertical direction.  Right: the whole view around the cone.
}
\label{Fig:Taylor}
}
%

We have two examples of conic D-branes, which mimic the Taylor cones. The first 
one is a D2-D0 bound state under a constant flux of a Ramond-Ramond (RR)
one-form field.
The D0-brane charges on the D2-brane are pulled by the flux, as in the Taylor cone.
We solve the DBI equation of motion and found a conic solution. The second example is 
a D$p$-brane under a constant flux of Neveu-Schwarz Neveu-Schwarz (NSNS)
two-form field.
Induced fundamental string charges on the D$p$-brane are pulled by the flux, and the D$p$-brane forms a conic shape. 

We can explicitly show the force balance on the conic D-branes, for the two examples as well as the previously known critical embedding of the probe D-brane with  
electric and magnetic fields in AdS/CFT.  The force balance is quite simple: the external force has two components, one for the direction parallel to the generating line of the cone and the other for perpendicular direction. The parallel force is canceled by the tension of the D-brane, while 
the perpendicular force is canceled by the surface tension of the round shape  of the surface, defined by an extrinsic curvature. 

Together with the example of the critical embedding of the probe D-brane in the electric field,
we find that all examples share the same property: universal cone angle. It is given by
\begin{eqnarray}
\theta_{\rm cone}=\arctan\sqrt{2(d_{\rm cone}-1)}\, ,
\label{coneangle1}
\end{eqnarray} 
where $d_{\rm cone}$ is the spatial dimensions of the cone (including the direction of the
generating line of the cone), {\it i.e.}, the cone is locally 
${\bf R}^{+} \times {\bf S}^{d_{\rm cone}-1}$. The ``quantized'' universal angle of 
D-brane cones is 
reminiscent of the Taylor cone angle.

The conic shape of the probe D-brane in the AdS/CFT example is closely related to
a phase transition. The critical embedding showing the cone appears at the phase boundary of the meson melting transition \cite{Erdmenger:2007bn,Albash:2007bq} 
caused by the electric field in supersymmetric QCD. The D-brane configuration
is decomposed into radial modes which correspond to meson expectation values,
and was found to exhibit a power-law spectrum \cite{Hashimoto:2014xta,
Hashimoto:2014dda,Hashimoto:2015psa}. 
Time-dependent simulations such as dynamically applied electric fields were performed
in \cite{Hashimoto:2014xta,Hashimoto:2014dda,Hashimoto:2014yza,Ishii:2014paa,Ishii:2015wua} and 
a similar power-law spectrum appears also there
\cite{Hashimoto:2014xta,
Hashimoto:2014dda, Ishii:2015wua}, where a
 singularity formation
resembles the famous Bizon-Rostworowski conjecture about AdS turbulent instability
\cite{Bizon:2011gg}. 
All the above turbulent behavior with a power-law spectrum was
confirmed numerically in the both static and time-dependent cases,
but in this paper 
we provide an analytical proof of 
such power law behavior
of the static ``turbulent meson condensation'', based on the conical shape of
the probe D-brane.

The organization of this paper is as follows. First in Sec.~\ref{Sec:balance},
we provide a generic analysis of a membrane cone under an external force,
for given stress-energy tensor on the membrane. Then in Sec.~3, we provide
three examples of conic D-brane solutions: a D2-D0 bound state in RR flux,
a D$p$-brane in NSNS flux, and a probe D7-brane in AdS/CFT. The first and the second examples are new. Then we check the force balance for all the cases, 
and find that the apex angle is universally given by (\ref{coneangle1}).
In Sec.~4, we give an analytic proof of the power law for static meson turbulence
on the D7-brane in AdS/CFT. The final section is for a conclusion and discussions.

\section{Membrane cone under a uniform external force}
\label{Sec:balance}

In this section, we summarize how a membrane can form a cone as a result
of a force balance on the cone surface. The external force has a component
perpendicular to the cone surface, which is canceled by the stress of the
membrane with the extrinsic curvature as for the round shape. First, we provide
an intuitive picture of the force balance, and then we provide a covariant
formulation.
Both lead to a specific formula of a universal angle at the apex of the cone.

\subsection{Force balance in Newtonian mechanics}

%
\FIGURE{ 
\includegraphics[width=7.7cm]{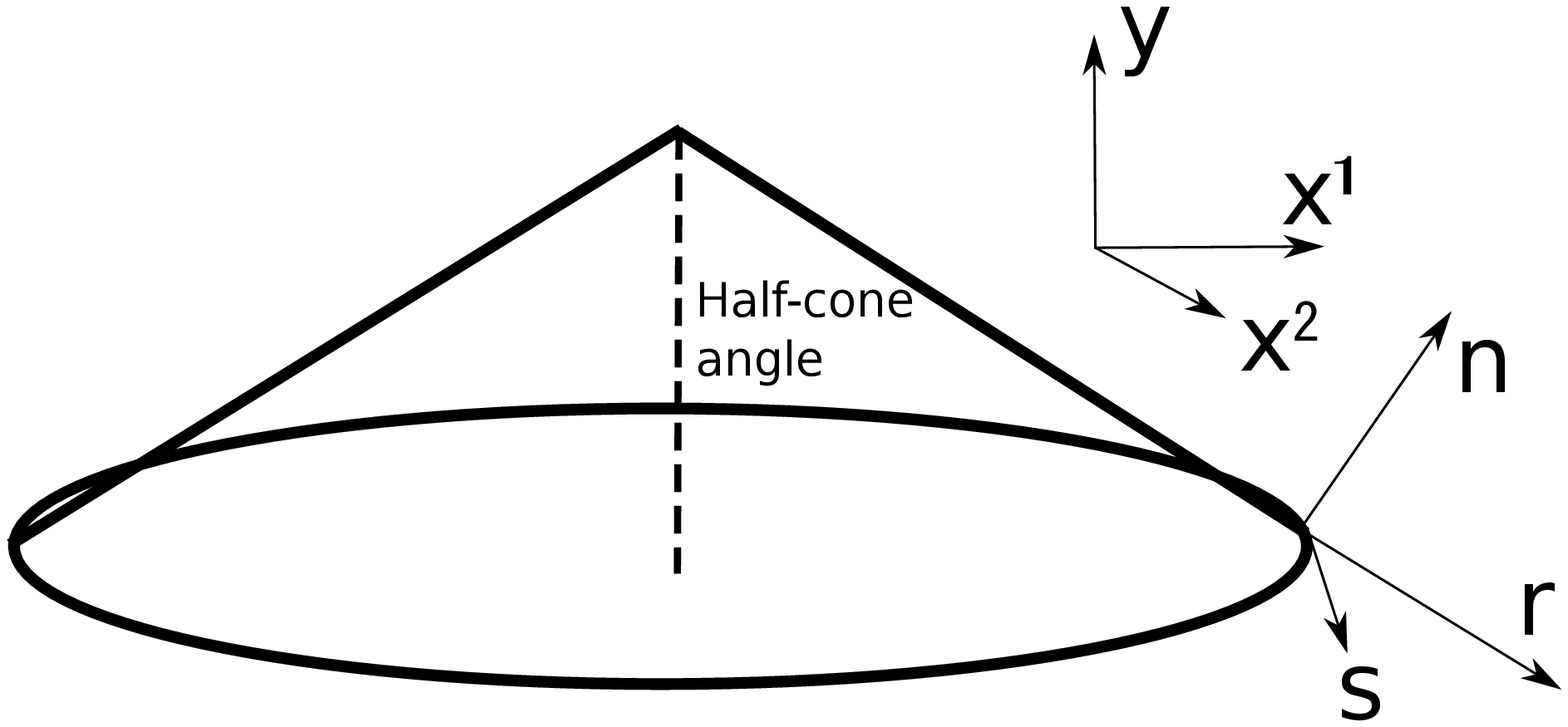}
\caption{The cone and our coordinate system. The cone axis
is taken along $y$, and the Half-cone angle $\theta_{\rm cone}$ 
is shown. On the surface, we have two unit tangent vectors $\vec{r}$
and $\vec{s}$ which are along radial / angular directions respectively,
and a normal vector $\vec{n}$.}
\label{Fig:conetangent}
}
%

Consider a membrane in a flat space with 3 spatial dimensions.
Let us assume that the system, including the external force, is
axially symmetric around the $x^3(=y)$ axis. 
Then the membrane configuration is given by 
\begin{eqnarray}
(x^1,x^2,y) = (x^1,x^2,\phi(\rho))
\end{eqnarray}
where $\rho \equiv \sqrt{(x^1)^2+(x^2)^2}$. Forming a cone means
the constraint 
\begin{eqnarray}
\frac{d \phi(\rho)}{d\rho} = \cot\theta_{\rm cone}
\end{eqnarray}
where $\theta_{\rm cone}$ is a constant, which is the half cone angle
(see Fig.~\ref{Fig:conetangent}).
Along the radial direction of the cone, we can define a radial coordinate 
\begin{eqnarray}
r = \frac{\rho}{\sin\theta_{\rm cone}}.
\end{eqnarray}

Let us consider a force balance condition.
The first force balance condition is along the cone surface.
The external force $F_r$ along the cone radial direction 
needs to be balanced by the surface tension as 
\begin{eqnarray}
|F_r| = -\partial_r T_{rr}
\label{Fr}
\end{eqnarray}
where $T_{rr}$ is the proper $rr$ component of the 
stress tensor on the conic membrane.
This force, similar to a hydrodynamic equilibrium, is given by a gradient of
the stress tensor.

The second force balance condition is along the direction perpendicular
to the cone surface. 
There is a surface stress tension caused by the curvature of
the membrane surface. The cone curvature is nontrivial along 
the circular direction of the cone. Calling 
the proper component of the stress tensor
along the circular direction (which we call $s$) on a point of the cone surface
as $T_{ss}$, then the Young-Laplace equation tells us the force
oriented inward the cone (perpendicular to the cone surface) is
balanced with the normal component of the external force as 
\begin{eqnarray}
|F_n| = -T_{ss}\frac{1}{r\tan\theta_{\rm cone}}.
\label{Fn}
\end{eqnarray}
In this force balance condition, 
the last factor is due to the curvature. 
In summary, we have two force balance conditions, 
(\ref{Fr}) and (\ref{Fn}). See Fig.~\ref{Fig:balance}.

%
\FIGURE{ 
\includegraphics[width=7cm]{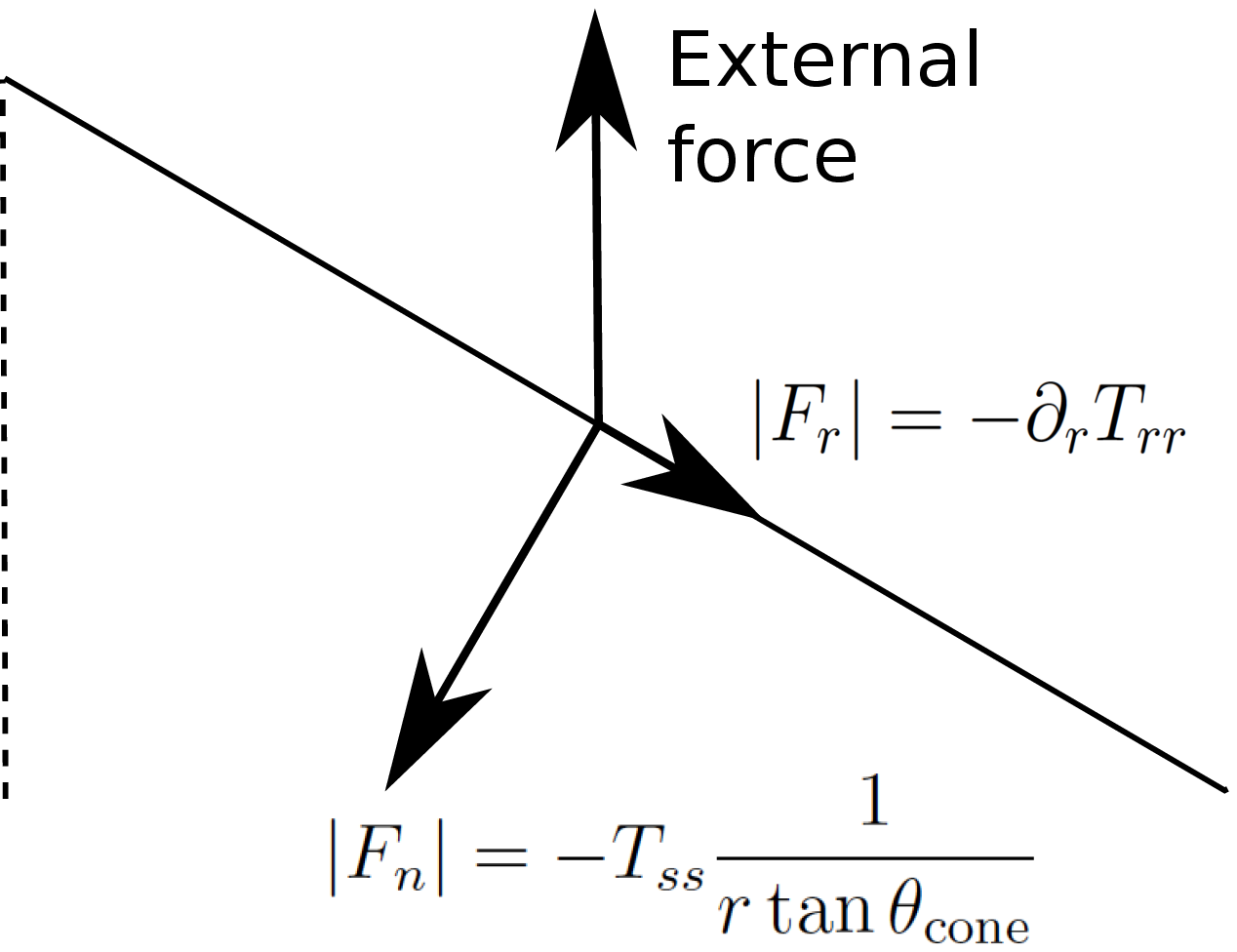}
\caption{The force balance at the surface.}
\label{Fig:balance}
}

It is important to note that the external force can be eliminated in
the force balance conditions (\ref{Fr}) and (\ref{Fn}), if
the combined vector
$\vec{F_r}+\vec{F_n}$  is oriented along $y$ axis.
This is a natural assumption, since 
the cone isometry prefers the symmetry of the external force configuration. 
For a given cone angle $\theta_{\rm cone}$,  
the total external force directed along $y$ axis means
\begin{eqnarray}
|F_n| = |F_r| \tan\theta_{\rm cone}.
\end{eqnarray}
Substituting (\ref{Fr}) and (\ref{Fn}) into this relation, 
we can eliminate the strength of the external force from 
the force balance conditions (\ref{Fr}) and (\ref{Fn}) as 
\begin{eqnarray}
T_{ss}= (\tan\theta_{\rm cone})^2 \, r  \frac{d}{dr} T_{rr} \, .
\label{fb}
\end{eqnarray}
Whenever this condition is satisfied for a diagonal stress tensor
which depends only on $r$, the force balance for the uniform force
directed along the axis of the cone is ensured.

Now, let us assume the membrane has only isotropic stress as 
\begin{eqnarray}
T_{ss}= T_{rr} \, .
\end{eqnarray}
The stress tensor can be approximated as a power
function near the apex of the cone $r\sim 0$, thus we simply assume
\begin{eqnarray}
T_{ss}= T_{rr} \simeq A r^\alpha \quad (r \sim 0)
\label{assumealpha}
\end{eqnarray}
where $A$ and $\alpha$ are constant parameters. Then the 
relation of the stress tensor components 
(\ref{fb}) can be easily solved as
\begin{eqnarray}
\theta_{\rm cone} = \arctan \sqrt{\frac{1}{\alpha}}.
\label{caa}
\end{eqnarray}
Therefore, the membrane dynamics determines completely the 
cone angle $\theta_{\rm cone}$, and furthermore,
the cone angle depends only on how the stress tensor behaves
near the apex, (\ref{assumealpha}).

Generically the energy momentum tensor on the membrane depends on 
properties of the material such as what kind of equation of state
the membrane has,  
how the membrane behaves under the external force, and so on. 
For example,
the charge distribution on the membrane is essential for the energy momentum
tensor, while the distribution normally depends on the binding energy and
the repulsive forces between the charges. 

Let us summarize what we have described here. We have assumed the following three things: (1) The system is axially symmetric in 3 spatial dimensions, 
(2) The force is along the axis of the symmetry,
and (3) the stress on the brane is isotropic. Then it follows that the
cone angle $\theta_{\rm cone}$ is given universally as (\ref{caa}).
It depends only on the radial power $\alpha$ 
 of the stress (\ref{assumealpha}). In particular, when the cone is formed, 
the stress at the cone apex vanishes. In other words, when the cone apex stress
does not vanish, any cone is not formed.


  \subsection{Covariant treatment}
  \label{sec:covariant}

  In this subsection, we shall provide a covariant formulation for
  dynamics of membranes in general curved spacetimes 
  to see force balance for various conic membranes.
  Let $\{x^\mu\}$ and $\{y^a\}$ be coordinates on the bulk spacetime and
  the worldvolume of a membrane, respectively.
  When the embedding of the membrane in the bulk spacetime is determined
  by $x^\mu = X^\mu(y^a)$, the induced metric on the membrane is given
  by 
  \begin{equation}
   h_{ab} = g_{\mu\nu} \partial_a X^\mu \partial_b X^\nu ,
  \end{equation}
  where $g_{\mu\nu}$ denotes the bulk spacetime metric.
  For later convenience, we define a projection tensor mapping from 
  vectors in the bulk to vectors on the membrane as
  $h_a{}^\mu \equiv \partial_a X^\mu$.

  It has been known that extrinsic and intrinsic 
  dynamics of the membrane are governed by the
  equations simplified geometrically as 
  \begin{align}
   T^{ab}K^\mu{}_{ab} =& - \mathcal{F}_\mathrm{n}^\mu , 
   \label{eq:extrinsic_eom}\\
   D_a T^{ab} =& \mathcal{F}_\mathrm{t}^b ,
   \label{eq:intrinsic_eom}
  \end{align}
  where $T^{ab}$ is the stress-energy tensor of the membrane and $D_a$
  denotes the covariant derivative with respect to the induced metric $h_{ab}$.
  $K^\mu{}_{ab}$ is
  the extrinsic curvature, which is given by%
  \footnote{
  Since $\partial_a = h_a{}^\mu \partial_\mu$ are coordinate basis on the
  submanifold representing the membrane, 
  $\partial_a$ and $\partial_b$ commute.
  It means  
  $[\partial_a, \partial_b]^\mu = h_a{}^\nu \nabla_\nu h_b{}^\mu -  
  h_b{}^\nu \nabla_\nu h_a{}^\mu = 0$
  and $K^\mu{}_{ab}$ is symmetric in $a$ and $b$.
  }
  \begin{equation}
  \begin{aligned}
   - K^\mu {}_{ab} \equiv& (g^\mu{}_\lambda - h^\mu{}_\lambda)
   h_a{}^\nu \nabla_\nu h_b{}^\lambda
   \\
   =& D_a D_b X^\mu + \Gamma^\mu{}_{\alpha\beta}
   D_a X^\alpha D_b X^\beta ,
  \end{aligned}
  \end{equation}
  where 
  $\Gamma^\mu{}_{\alpha\beta}$ is the Christoffel symbol for the
  bulk metric.
  $\mathcal{F}_\mathrm{n}^\mu$ and $\mathcal{F}_\mathrm{t}^a$ are
  normal and tangential components of external forces acting on some
  charges with the membrane, which are defined by 
  $\mathcal{F}^\mu = \mathcal{F}_\mathrm{n}^\mu
  + \mathcal{F}_\mathrm{t}^a h_a{}^\mu$ and 
  $g_{\mu\nu} h_a{}^\mu \mathcal{F}_\mathrm{n}^\nu = 0$.
  We note that 
  Eq.~(\ref{eq:extrinsic_eom}) corresponds to Eq.~(\ref{Fn}) in the previous
  subsection and Eq.~(\ref{eq:intrinsic_eom}) corresponds to Eq.~(\ref{Fr}),
  which yields fundamental equations of hydrodynamics or elastic dynamics.

  Now, we suppose that an ``axisymmetric'' bulk spacetime is given by 
  \begin{equation}
   \begin{aligned}
   g_{\mu\nu} dx^\mu dx^\nu =& 
    \mathsf{A}_{ij}(\rho, \zeta)dy^i dy^j
    + \mathsf{B}(\rho,\zeta)(d\rho^2 + d\zeta^2)
    + \mathsf{C}(\rho,\zeta)d\Omega^2_{d-1} \\
    & + \mathsf{D}_{kl}(\rho,\zeta)dw^k dw^l ,
   \end{aligned}
  \end{equation}
  where $\rho$ and $\zeta$ can be interpreted as radial and horizontal
  coordinates in a usual cylindrical coordinate system.
  Note that we assume 
\begin{eqnarray}
  \partial \mathsf{A}_{ij}/\partial\rho|_{\rho=0} = 0, \quad 
  (2\sqrt{\mathsf{BC}})^{-1} \partial\mathsf{C}/\partial\rho|_{\rho=0} = 1, \quad
  \mathsf{C}|_{\rho=0} = 0
  \label{assumeregular}
  \end{eqnarray}
  for the bulk spacetime to be regular at the axis $\rho=0$.
  When a membrane is axisymmetrically embedded by functions 
  $\zeta = \phi(\rho)$ and $w^k = \text{const.}$, 
  the induced metric on the membrane becomes 
  \begin{equation}
   \begin{aligned}
    h_{ab} dy^a dy^b =&
    \mathsf{A}_{ij}(\rho, \phi(\rho)) dy^i dy^j
    + [1+\phi'(\rho)^2] \mathsf{B}(\rho,\phi(\rho))d\rho^2 \\
    &~ + \mathsf{C}(\rho,\phi(\rho))d\Omega^2_{d-1} .    
   \end{aligned}
  \end{equation}

  By introducing $\sin\theta(\rho) \equiv 1/\sqrt{1+\phi'^2}$ and 
  $\cos\theta(\rho) \equiv \phi'/\sqrt{1+\phi'^2}$, 
  the unit normal one-form and vector along the non-trivial normal
  direction for the membrane are  
  \begin{equation}
   n_\mu dx^\mu = 
    \sqrt{\mathsf{B}}
    (\cos\theta d\rho - \sin\theta d\zeta) ,\quad
    n^\mu \partial_\mu = 
    \frac{1}{\sqrt{\mathsf{B}}}
    (\cos\theta\partial_\rho - \sin\theta\partial_\zeta) .
  \end{equation}
  On the other hand, the unit tangent one-form and vector along the radial direction 
  (namely the generatrix of the cone) are 
  \begin{equation}
   r_\mu dx^\mu = 
    \sqrt{\mathsf{B}}
    (\sin\theta d\rho + \cos\theta d\zeta) ,\quad
    r^\mu \partial_\mu = 
    \frac{1}{\sqrt{\mathsf{B}}}
    (\sin\theta\partial_\rho + \cos\theta\partial_\zeta) .
  \end{equation}
  Note that on the membrane these one-form and vector can be written as 
  $r_a dy^a = \sqrt{\mathsf{B}}/\sin\theta d\rho$ and 
  $r^a\partial_a = \sin\theta/\sqrt{\mathsf{B}} \partial_\rho$.

  We assume that the stress-energy tensor of the membrane has the
  following form: 
  \begin{equation}
   T_{ab} = \tau_{ab} - \sigma (r_a r_b + s_{ab}) ,
   \label{Ttausigma}
  \end{equation}
  where 
  $s_{ab}$ is the spherical part of the induced metric, that is 
  $s_{ab}dy^ady^b = \mathsf{C}d\Omega^2_{d-1}$.
  Since $\tau_{ab}$ are components other than those on the cone, 
  $\tau_{ab}r^b = 0$ and $\tau_{ab}s^{bc} = 0$ satisfy.
  If $\sigma > 0$, it means that the membrane has isotropic tension on
  the cone.\footnote{
  Note that for Nambu-Goto branes the energy density is equal to the tension, 
  but in general they are different. In appendix A, we study the stress-energy tensor for a system described by Dirac-Born-Infeld (DBI) action (see (\ref{eq:DEF_energy_momentum})).
  }

  For the normal direction along $n^\mu$, 
  from Eq.~(\ref{eq:extrinsic_eom}) we have 
  \begin{equation}
   \begin{aligned}
    \mathcal{F}^\mu n_{\mu} =&
    - T^{ab} K_{ab} \\
    =& - \tau^{ab} K_{ab} + \sigma K_{ab}r^a r^b + \sigma s^{ab}K_{ab} ,
   \end{aligned}
   \label{eq:normal_F}
  \end{equation}
  where $n_\mu K^\mu{}_{ab} = K_{ab} \equiv h_a{}^\mu h_b{}^\nu
  \nabla_\mu n_\nu$.
  For the tangential direction along $r^\mu$, 
  from Eq.~(\ref{eq:intrinsic_eom}) we have 
  \begin{equation}
   \begin{aligned}
    \mathcal{F}^\mu r_{\mu} =& r_b D_a T^{ab} \\
    =& - D_a(\sigma r^a) + \sigma s^{ab}D_a r_b - \tau^{ab}D_a r_b
   \end{aligned}
   \label{eq:radial_F}
  \end{equation} 

  If the external force is along the axis of the cone, namely
  only $\mathcal{F}^\zeta$ is a non-zero component%
  \footnote{
  Here, we consider both of the external force and the normal vector
  are in the direction of the same side for the membrane, namely
  $\mathcal{F}^\mu n_\mu > 0$.
  }, 
  then   
  we obtain 
  $\mathcal{F}^\mu (r_\mu \sin\theta + n_\mu \cos\theta) = 0$. 
  Combining Eqs.~(\ref{eq:normal_F}) and (\ref{eq:radial_F}) yields 
  \begin{equation}
   - \sin\theta D_a(\sigma r^a)
    + \sigma s^{ab} \frac{1}{2\sqrt{\mathsf{B}}} \partial_\rho s_{ab}
    - \tau^{ab} \frac{1}{2\sqrt{\mathsf{B}}} \partial_\rho \mathsf{A}_{ab}
    + \cos\theta
    \sigma r^ar^bK_{ab} = 0 ,
  \end{equation}
  where we have used 
  \begin{equation}
   \frac{1}{\sqrt{\mathsf{B}}} \frac{\partial}{\partial\rho}
    = \sin\theta r^\mu \partial_\mu + \cos\theta n^\mu \partial_\mu.
  \end{equation}
  By using the metric functions explicitly, it can be written as   
  \begin{equation}
   \begin{aligned}
    \frac{\sin\theta}{\sqrt{-\mathsf{AB}}}
    \frac{d}{d\rho}(\sqrt{\mathsf{-A}} \sigma \sin\theta)
    - \sigma \cos\theta n^\mu \partial_\mu 
    \log(\sqrt{\mathsf{B}} \mathsf{C}^{(d-1)/2})
    + \frac{1}{2\sqrt{\mathsf{B}}}\tau^{ij} \partial_\rho \mathsf{A}_{ij}
    = 0 ,
   \end{aligned}
   \label{eq:covariant_balance_condition}
  \end{equation}
  where $\mathsf{A} \equiv \det \mathsf{A}_{ij}$.
  
  Since we have assumed that the spacetime is regular at $\rho = 0$ 
  by imposing the regularity condition (\ref{assumeregular}), 
  the metric function can be expanded around $\rho=0$ as 
  \begin{equation}
   \mathsf{A}_{ij} = \mathsf{A}_{0ij}(\zeta) + \mathcal{O}(\rho^2), \quad
    \mathsf{B} = \mathsf{B}_0(\zeta) + \mathcal{O}(\rho^2), \quad
    \mathsf{C} = \mathsf{B}_0(\zeta) \rho^2 + \mathcal{O}(\rho^4) , 
  \end{equation}
  where a regular polar coordinate needs $\mathsf{B}_0 \neq 0$.
  In addition, we assume 
  $\det \mathsf{A}_{0ij} \neq 0$ on the membrane , which means 
  that the membrane does not touch Killing horizons in the bulk.%
  \footnote{
  On the other hand, if the membrane touches
  bulk black holes, such as the
  critical embedding at thermal phase transition,
  $\sqrt{-\mathsf{A}}\sim 0$ in
  Eq.~(\ref{eq:covariant_balance_condition}) plays a significant role.
  Physically, this means that infinite gravitational blue-shift near the horizon
  becomes so significant rather than the matter distribution.
  }
  If we 
  assume in (\ref{Ttausigma}) that the tension $\sigma$ plays a dominant role, 
  $\sigma \gg |\tau| \rho^2$  for $\rho\sim 0$, 
  we have the following condition,  
  \begin{equation}
   (\sin\theta_\mathrm{cone})^2 \frac{d\sigma}{d\rho} 
    \simeq (d-1) (\cos\theta_\mathrm{cone})^2 \frac{\sigma}{\rho} ,
  \end{equation}
  where $\theta_\mathrm{cone} \equiv \theta(\rho=0)$.
  Now, suppose the tension behaves as $\sigma \sim \rho^\alpha$ 
  ($\alpha > 0$) near the apex of cone $\rho \sim 0$, 
  then the angle of the cone becomes 
  \begin{equation}
   \theta_\mathrm{cone} = \arctan \sqrt{\frac{d-1}{\alpha}} .
    \label{eq:general_formula}
  \end{equation}
  This is our formula for the cone angle, simply written only by the cone dimension
  and the scaling of the tension.

Obviously the cone angle formula (\ref{eq:general_formula}) generalizes the
previous formula (\ref{caa}) in the following respect: (1) it works in a general geometry, 
(2) it uses generic energy momentum tensor on the brane, and (3) the brane can have arbitrary worldvolume dimensions. In the next section, we study examples of
D-branes in string theory, and will find $\alpha=1/2$ for every example we consider.


\section{Conic D-branes and universal cone angle }

The phenomenon of the Taylor cones suggest that D-branes in superstring theory
can develop a conic shape under some background field. Furthermore, it suggests
the existence of a universal cone angle, as the Taylor cones are formed by
quite a simple mechanism which can be generalized to higher dimensions.

In this section, we provide three new examples of conic D-branes (Sec.~\ref{sec:exRR}, \ref{sec:exNSNS}, \ref{sec:exAdS}), and
provide a universal formula for the cone angle which just depends on the 
dimensions of the cone worldvolume (Sec.~\ref{cone-angle}).

The Taylor cone is formed simply because ion charges on the liquid surface
are pulled by a background electric field and cancel the liquid surface tension.
So it is natural to expect higher dimensional analogues in string theory.
The three examples of the background field 
we present here are: The case of Ramond-Ramond (RR)
flux, the case of Neveu-Schwarz Neveu-Schwarz (NSNS) flux and 
the case of AdS/CFT with electric and magnetic fields.


\subsection{D-brane cone in RR flux}
\label{sec:exRR}

\subsubsection{Conic solution}

The first example is a D-brane in a constant RR flux background in a flat spacetime.
The D-brane is an analogue of the membrane-like surface of the Taylor cones.
To put ``ions'' on the membrane, we consider lower dimensional D-branes on the D-brane. 
To have a bound state, it is suitable to choose D$(p-2)$-branes bound on a single
D$p$-brane. Then we turn on a background RR flux which pulls the bound D$(p-2)$-branes. 
The RR gauge field should be $(p-1)$-form field for which the D$(p-2)$-branes
are charged. Below we shall show that conic shape of the D$p$-brane can be obtained
as a classical solution of the D$p$-brane worldvolume effective action.

We can choose $p=2$ without loosing generality, as all other choices of $p$
are obtained by T-dualities. The bound D$0$-branes are described by a
field strength of the D-brane gauge field. The D2-brane effective action is
a Dirac-Born-Infeld (DBI) action plus a coupling to the RR field,
\begin{eqnarray}
S &=& -{\cal T}_2 \int dx^0 dx^1 dx^2 \; 
\sqrt{-\det (\eta_{ab} + 2\pi\alpha' F_{ab} + \partial_a \phi \partial_b \phi)}
\nonumber \\
&&{} - \frac{{\cal T}_2}{2} \int dx^0 dx^1 dx^2 \; 
 2\pi\alpha' F_{ab} C_c \epsilon^{abc} \, .
\end{eqnarray}
Here $a,b=0,1,2$ are the worldvolume directions of the D2-brane, and $F_{ab}$
is the gauge field strength on the D2-brane. $C_c$ is the RR one-form in
the bulk space. $\phi$ is a scalar field on the D-brane which measures the
displacement of the position of the D2-brane in the direction transverse to the
D2-brane worldvolume. We chose one direction $\phi$ 
among 7 transverse directions
for simplicity.

We turn on a temporal component of the RR background
\begin{eqnarray}
C_0 = C_0[\phi]
\end{eqnarray}
where $C_0[\phi]$ is an arbitrary functional of $\phi(x)$. Having this is
indeed equivalent to have a nontrivial background RR field strength
$H_{\phi 0}^{\rm RR}\equiv \partial_\phi C_0[\phi]$. 

We are interested in a static D2-brane configuration without any electric field on it,
so the action can be simplified\footnote{The equation of motion for $A_0$ (which is Gauss law) is satisfied by $F_{01}=F_{02}=\partial_0 \phi=0$ because the action is quadratic in
these components.} with $F_{01}=F_{02}=\partial_0 \phi=0$. 
We obtain the action as
\begin{eqnarray}
S = {\rm const.}\int \! dx^0 dx^1 dx^2 \;
\left[
-\sqrt{1+ (2\pi\alpha'F_{12})^2 + (\partial_1\phi)^2 + (\partial_2 \phi)^2}
+ {2\pi\alpha'} C_0[\phi] F_{12} \right] \, .
\end{eqnarray}
Here the overall normalization of the RR field $C_0$ was chosen
to simplify the Lagrangian.%
\footnote{$\epsilon_{012}=+1$ and $\epsilon^{012} = - 1$ in our convention.}

The equations of motion are
\begin{eqnarray}
&&
\partial_i \left[\frac{\partial_i \phi}{\sqrt{1+ (2\pi\alpha'F_{12})^2 +
	    (\partial_1\phi)^2 + (\partial_2 \phi)^2}}\right]
+ {2\pi\alpha'} \frac{dC_0[\phi]}{d\phi} F_{12} = 0 \, ,
\label{phieq}
\\
&&
\partial_i \left[\frac{2\pi\alpha'F_{12}}{\sqrt{1+ (2\pi\alpha'F_{12})^2 + (\partial_1\phi)^2 + (\partial_2 \phi)^2}}\right] 
- \partial_i C_0[\phi] = 0 \, ,
\end{eqnarray}
where $i=1,2$.
The second equation can be integrated to give 
\begin{eqnarray}
2\pi\alpha'F_{12} = C_0[\phi] \sqrt{\frac{1 + (\partial_i \phi)^2}{1-(C_0[\phi])^2}} \, .
\label{F12sol}
\end{eqnarray}
Here we absorbed the integration constant to a redefinition of $C_0[\phi]$ 
by a constant shift
($C_0[\phi] \rightarrow C_0[\phi] +{\rm const.}$) without losing the generality, and chose a sign for our later convenience.
It is amusing that (\ref{F12sol}) is a first order differential equation, so we may call 
(\ref{F12sol})
a ``BPS'' equation. 

Substituting (\ref{F12sol}) to (\ref{phieq}), we
obtain a differential equation for $\phi$,
\begin{eqnarray}
\partial_i \left[
\frac{\partial_i \phi}{\sqrt{1 + (\partial_j \phi)^2}} \sqrt{1-(C_0[\phi])^2}
\right]
+ \frac{d C_0[\phi]}{d\phi} C_0[\phi]  
\sqrt{\frac{1+(\partial_j \phi)^2}{1-(C_0[\phi])^2}} = 0 \, .
\label{fullphieom}
\end{eqnarray}
The equation is singular when $C_0[\phi]=\pm1$. 
In fact, we will see that this 
point in the bulk provides the tip of the cone solution. 
It is instructive to note that this equation (\ref{fullphieom}) can
be derived from an ``effective'' action
\begin{eqnarray}
S_{\rm cone} = \int \! dx^1 dx^2 \;
\sqrt{(1+(\partial_i \phi)^2)(1-(C_0[\phi])^2)}
\end{eqnarray} 

We consider a typical RR background, that is, a 
constant RR field strength along the direction $\phi$, as
\begin{eqnarray}
\partial_\phi C_0 = c
\end{eqnarray}
where $c (>0)$ is a constant parameter.
Up to a gauge choice we can take
\begin{eqnarray}
C_0 =  c \, \phi \, .
\end{eqnarray}
This background can be thought of as a local approximation of generic $C_0[\phi]$
background. In fact, to show the existence of conic D-branes the local approximation
is enough.
For this constant RR flux background, the singularity 
in the equation (\ref{fullphieom}) is found at $\phi = 1/c$. 
We assume the rotational symmetry for solutions, and 
expand $\phi$ around this singular value,
\begin{eqnarray}
\phi = 1/c - f(\rho)
\end{eqnarray}
where $\rho \equiv \sqrt{(x^1)^2+(x^2)^2}$ is the radial coordinate on the D2-brane
worldvolume.
Then, to the leading order in $f(\rho)$, we obtain an equation for $f(\rho)$ as 
\begin{eqnarray}
\frac{1}{\rho} 
\partial_\rho
\left[
\rho \frac{\partial_\rho f}{\sqrt{1 + (\partial_\rho f)^2}} \sqrt{f}
\right]
- \frac{1}{2\sqrt{f}}\sqrt{1 + (\partial_\rho f)^2} = 0\, .
\label{feom}
\end{eqnarray}
Again, 
the equation can be obtained from an ``effective'' one-dimensional action
\begin{eqnarray}
S_{\rm cone} = \int \! d\rho \; \rho \sqrt{f(\rho) \, \left(1 + \left(\partial_\rho f(\rho)\right)^2\right)} \, .
\label{effectiveL}
\end{eqnarray}
We will find later that this effective 
action can be universally found in string theory.

%
\FIGURE{ 
\includegraphics[width=8cm]{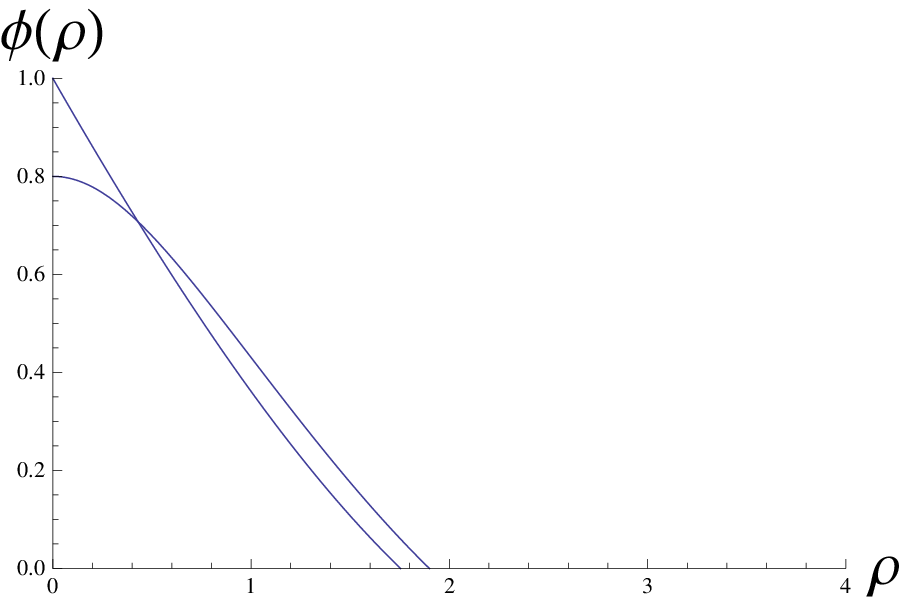}
\caption{A numerical solution $\phi(\rho)$
for the full equation of motion (\ref{fullphieom}) with $C_0[\phi]= c\phi$.
The straight line is for $\phi(0)=1/c$ which is the ``singular'' point, 
and a curved line is for $\phi(0)=4/5c$ which is below the singular point. 
We took $c=1$ for this plot. 
We find that
a cone is formed for the case $\phi(\rho=0)=1/c$.}
\label{Fig:cone}
}
%
To solve the equation (\ref{feom}), we consider the following ansatz
\begin{eqnarray}
\phi = \frac{1}{c} - a \rho^b \quad (b>0) \, ,
\end{eqnarray}
which goes to $\phi=1/c$ at $\rho=0$. 
Substituting this to (\ref{feom}),
we can show that any nontrivial solution has a unique form
\begin{eqnarray}
a = \frac{1}{\sqrt{2}} \, , \quad b=1 \, .
\label{uniqueab}
\end{eqnarray}
The value $b=1$ indeed shows a cone, as the radius of the section of the D2-brane at
fixed $\phi$ is given by a linear function of $\rho$. So, finally we could show that
the D2-brane configuration reaching $\phi=1/c$ is conic: 
\begin{eqnarray}
\phi = \frac{1}{c} - \frac{1}{\sqrt{2}} \rho  + \mbox{higher}\, .
\label{coneRRsol}
\end{eqnarray}

We have two comments on the conic D-brane configuration. First, let us evaluate the D0-brane charge density. Substituting the near-tip solution to (\ref{F12sol}), we find
\begin{eqnarray}
2\pi\alpha'F_{12} = 3^{1/2} 2^{-3/4} c^{-1/2} \frac{1}{\sqrt{\rho}}
\end{eqnarray}
near $\rho=0$. This shows that the bound D0-brane charge density is proportional 
to $1/\sqrt{l}$ where $l$ is the distance from the tip of the cone. 
On the other hand, it is known that Taylor cone has a charge distribution
$\sim 1/\sqrt{l}$ around the tip of the cone. So, our result is 
similar to 
the Taylor cone.

The second comment is about the asymptotics. Solutions of the full equation of
motion (\ref{fullphieom}) for $C_0[\phi]=c \phi$ 
are shown in Fig.\ref{Fig:cone}. They indicate
that even at large $\rho$, the shape follows that of a cone. 
However, it is not that case. As seen from the equation (\ref{fullphieom}),
the configuration of $\phi$ is limited to the region
\begin{eqnarray}
-\frac{1}{c} < \phi(\rho) < \frac{1}{c} \, .
\end{eqnarray}
Otherwise the equation (\ref{F12sol}) provides an imaginary
field strength $F_{12}$. So, the brane configuration 
terminates when it reaches $\phi = -1/c$. This strange behavior is due to
our approximation of constant RR field strength. Generically in string theory,
the RR field varies in space, and our constant approximation is valid only locally.
The asymptotics depends on a global structure of the RR background.

%
\FIGURE{ 
\includegraphics[width=11cm]{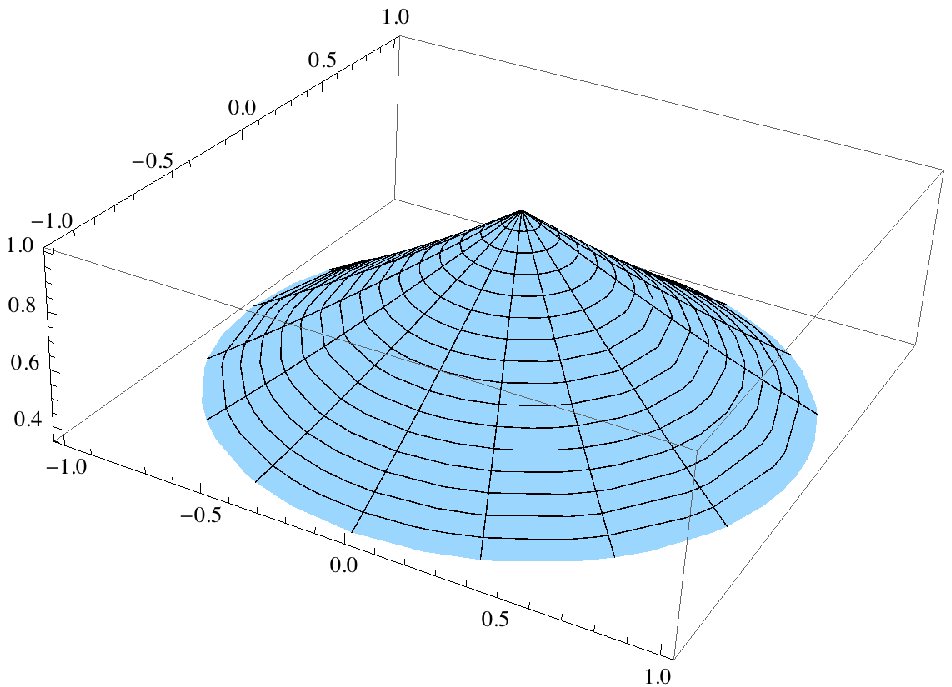}
\caption{A view of the conic D2-brane. The parameters are the one
in the previous figure. }
\label{Fig:coneRR}
}
%

\subsubsection{Force balance of the cone}

Let us follow the argument in Sec.~\ref{sec:covariant} about a covariant treatment of
the force balance, for this conic D-brane in a RR flux. We can explicitly see that
the force balance condition provides the conic solution (\ref{coneRRsol}).
   
   We consider the D$2$-brane described by the embedding function 
   $y=\phi(x^1, x^2)$ in the RR background $C_0 = cy$ in the flat
   spacetime 
   \begin{equation}
    g_{\mu\nu}dx^\mu dx^\nu 
     = - dt^2 + (dx^1)^2 + (dx^2)^2 + d\vec{y}^2_{d-3} .
   \end{equation} 
   Assuming static and axisymmetric, the induced metric on the D$2$-brane is 
   \begin{equation}
    h_{ab}dx^a dx^b = - dt^2 + (1+\phi'^2) d\rho^2
     + \rho^2 d\theta^2 ,
   \end{equation}
   where $\rho^2 = (x^1)^2 + (x^2)^2$.
   The unit vector and $1$-form normal to the brane are  
   \begin{equation}
    n^\mu \partial_\mu = \frac{1}{\sqrt{1+\phi'^2}}
     (\phi'\partial_\rho - \partial_y) ,\quad
     n_\mu dx^\mu = \frac{1}{\sqrt{1+\phi'^2}}
     (\phi' d\rho - dy) .
   \end{equation}
   The unit vector and $1$-form tangent to the brane are 
   \begin{equation}
    r^\mu \partial_\mu = \frac{1}{\sqrt{1+\phi'^2}}
     (\partial_\rho + \phi'\partial_y) ,\quad
     r_\mu dx^\mu = \frac{1}{\sqrt{1+\phi'^2}}
     (d\rho + \phi'dy) .
   \end{equation} 
   We note that, strictly speaking, since there are many codimensions,
   other directions normal to the brane exist.
   However, we can focus on only the above normal vector $n$ because for
   the other directions the brane is trivially embedded.
   Non-vanishing components of the extrinsic curvature 
   $n_\mu K^\mu{}_{ab} = K_{ab} \equiv \frac{1}{2}\mathcal{L}_n h_{ab}$ are 
   \begin{equation}
    K_{\mu\nu}r^\mu r^\nu 
     = \left(\frac{\phi'}{\sqrt{1+\phi'^2}}\right)' ,\quad 
    K^\theta{}_\theta = \frac{\phi'}{\sqrt{1+\phi'^2}} \frac{1}{\rho} .
   \end{equation}
   By using 
   \begin{equation}
    2\pi\alpha'F_{12} = C_0[\phi] \sqrt{\frac{1+\phi'^2}{1-(C_0[\phi])^2}}
   \end{equation}
   from Eq.~(\ref{F12sol}), 
   non-vanishing components of the stress-energy tensor are 
   \begin{equation}
    \begin{aligned}
       T^0{}_0 =& - \sqrt{\frac{1+\phi'^2 + (2\pi\alpha'F_{12})^2}{1+\phi'^2}}
     = - \frac{1}{\sqrt{1-(C_0[\phi])^2}}
     ,\\
     T^\theta{}_\theta =& T_{\mu\nu}r^\mu r^\nu 
     = - \sqrt{\frac{1+\phi'^2}{1+\phi'^2 + (2\pi\alpha'F_{12})^2}} 
     = - \sqrt{1-(C_0[\phi])^2} .
    \end{aligned}
   \end{equation}
   It turns out that the isotropic tension 
   $\sigma=-T^\theta{}_\theta=-T_{\mu\nu}r^\mu r^\nu$ defined in
   Eq.~(\ref{Ttausigma}) is now given by 
   $\sigma = \sqrt{1-(C_0[\phi])^2}$, which will vanish when 
   $C_0[\phi] = \pm 1$. 
   
   The equation of motion is alternatively written as
   \begin{equation}
    T_{\mu\nu}r^\mu r^\nu K_{\alpha\beta}r^\alpha r^\beta 
     + T^\theta{}_\theta K^\theta{}_\theta 
     = - n_\mu \mathcal{F}^\mu ,
   \end{equation}
   which is nothing but extrinsic force balance~(\ref{eq:extrinsic_eom}) in
   Sec.~\ref{sec:covariant}.
   In this case the external force is given by 
   $\mathcal F^\mu = - \frac{1}{2} F_{ab}e^{abc}H^\mu{}_c$ .
   If we suppose that when the tension vanishes at 
   $C_0[\phi]^2 = 1$ the brane becomes conical, namely $\rho = 0$, 
   the equation of motion reduces to 
   \begin{equation}
    n_\mu \mathcal{F}^\mu \simeq - T^\theta{}_\theta K^\theta{}_\theta .
   \end{equation}
   It can be rewritten as 
   \begin{equation}
    - \frac{1}{\sqrt{1+\phi'^2}} \frac{C_0}{\sqrt{1-C_0^2}} \frac{dC_0}{d\phi}
    \simeq
     \sqrt{1-C_0^2} \frac{\phi'}{\sqrt{1+\phi'^2}} \frac{1}{\rho} .
   \end{equation}

Solving the force balance equation tells us 
   \begin{equation}
    \phi'|_{\rho=0} = - 1/\sqrt{2} .
   \end{equation}
This is nothing but the conic solution (\ref{coneRRsol}). Here we have found that
the force balance condition is nothing but the equations of motion of the D-brane. In other words, solution of the equations of motion should satisfy the force balance condition necessarily.


\subsection{D-brane cone in NSNS flux}
\label{sec:exNSNS}

\subsubsection{Conic solution}

The next example is a D-brane in a NSNS flux background in a flat spacetime.
The charged bound object this time is fundamental strings, which couple
to the NSNS gauge field in the bulk. Interestingly, the NSNS flux in the background 
induces a fundamental string charge on the D-brane, so, this time we do not
need to prepare for the charged object on the D-brane from the first place,
as we will see below. The situation is contrary to the previous D2D0 case
where we needed $F_{12}$ on the D2-brane.

Let us consider a single D$p$-brane in a constant 
background NSNS flux $H_{01\phi}=c$. (The constancy is again only for simplicity,
and one can explore full configuration once the explicit NSNS flux is given.)
Here $\phi$ is a direction transverse to the D-brane, and $x^1$ is along the D-brane worldvolume. We shall show the existence of the cone configuration of the D-brane.

We choose a natural gauge for the NSNS flux, 
\begin{eqnarray}
B_{01} = c \phi \, , 
\label{B01}
\end{eqnarray}
and assume that
the D-brane shape depends only on the D-brane world volume scalar field
\begin{eqnarray}
\phi = \phi (\rho)
\end{eqnarray}
where $\rho$ is the radial coordinate on the D-brane worldvolume except for the 
$x^1$ direction, $\rho \equiv \sqrt{(x^2)^2 + \cdots + (x^{p})^2}$. 
The D-brane effective action is given by
\begin{eqnarray}
S &=& -{\cal T}_p \int \! dx^{p+1}x
\sqrt{-\det(\eta + B + \partial \phi \partial \phi )}  
\nonumber \\
& = & -{\cal T}_p \int \! dx^0 dx^1 d\rho \; V_{p-2} \rho^{p-2}\; 
\sqrt{(1+(d\phi/d\rho)^2) (1-B_{01}^2)}
\label{SDb}
\end{eqnarray}
where $V_{p-2}$ is the volume of the unit $(p-2)$-dimensional sphere.
Now, we substitute (\ref{B01}) and notice that the Lagrangian density
vanishes at the point $\phi = 1/c$ in the bulk spacetime. For the region
$\phi > 1/c$ in the bulk spacetime, the D-brane action becomes imaginary.

We shall show that, for the D-brane to reach this ``singular surface'' in the bulk,
the D-brane shape becomes conical. Let us expand the D-brane scalar field
around this singular surface as
\begin{eqnarray}
\phi (\rho) = \frac{1}{c}- f(\rho) \, ,
\end{eqnarray}
where $f(\rho=0)=0$. 
We are interested in the region close to the singular surface, so we assume 
$f(\rho)$ is small, and substitute it to the action (\ref{SDb}). 
Then, at the leading order in $f(\rho)$, we obtain 
\begin{eqnarray}
S = 
 -{\cal T}_p \int \! dx^0 dx^1 d\rho \; V_{p-2} \rho^{p-2}\; 
\sqrt{2f \, \left(1+(f')^2\right)} \, .
\label{f1ff}
\end{eqnarray}
The equation of motion for the scalar field is
\begin{eqnarray}
\frac{d}{d\rho}\left(
\rho^{p-2}\frac{f'}{\sqrt{1+(f')^2}}\sqrt{f}
\right)-\rho^{p-2} \frac{1}{2 \sqrt{f}}\sqrt{1+(f')^2} = 0 \, .
\label{eomf}
\end{eqnarray}
Near the origin $\rho=0$, the function $f(\rho)$ can be approximated by
\begin{eqnarray}
f (\rho) = a \rho^b\, , \quad a>0, \quad b>0 \, .
\end{eqnarray}
The equation of motion (\ref{eomf}) can be easily shown to have a nontrivial solution
only for $b=1$, and we find 
\begin{eqnarray}
a= \frac{1}{\sqrt{2p-4}}, \quad b=1\, .
\end{eqnarray}
This unique solution corresponds to the D-brane configuration near $\rho=0$, 
\begin{eqnarray}
\phi(\rho) = \frac{1}{c} - \frac{1}{\sqrt{2p-4}} \rho + \mbox{higher in $\rho$} \, .
\label{branecone}
\end{eqnarray}
We find that the D-brane forms a cone.

\subsubsection{Force balance of the cone}

Again, we shall investigate the force balance condition for this D-brane cone in
the NSNS flux, and will see that the condition is met with the equations of motion.

  We consider ($d+1$)-dimensional flat spacetime, 
  \begin{equation}
   \begin{aligned}
    g_{\mu\nu}dx^\mu dx^\nu =& 
    - dt^2 + d\vec{x}^2_p + d\vec{y}^2_{d-p} \\
    =& - dt^2 + dx_1^2 + d\rho^2 + \rho^2 d\Omega_{p-2}^2 + d\vec{y}^2_{d-p},
   \end{aligned}
  \end{equation}
  where $\rho^2 = x_2^2 + \cdots x_p^2$.
  Assuming that the NSNS field is 
  \begin{equation}
   B_{\mu\nu} dx^\mu \wedge dx^\nu = 2cy dt\wedge dx_1 , 
  \end{equation}
  we have the field strength  
  $H_{\mu\alpha\beta} = \partial_\mu B_{\alpha\beta} + \partial_\alpha
  B_{\beta\mu} + \partial_\beta B_{\mu\alpha}$ as  
  \begin{equation}
   H_{01y} = c .
  \end{equation}

  When the embedding function for the brane is given by $y=\phi(\rho)$,
  the induced metric is 
  \begin{equation}
   h_{ab}dx^a dx^b =
    - dt^2 + dx_1^2 + (1+\phi'^2) d\rho^2 + \rho^2 d\Omega^2_{p-2} , 
  \end{equation}
  and the effective metric 
  $\gamma_{ab} \equiv h_{ab} + B_{ac}B_{bd}h^{cd}$ is 
  \begin{equation}
   \gamma_{ab} dx^a dx^b =
    - (1-c^2\phi^2)dt^2 + (1-c^2\phi^2)dx_1^2
    + (1+\phi'^2) d\rho^2 + \rho^2 d\Omega^2_{p-2} .
  \end{equation}
  The unit vector and $1$-form normal to the brane are 
  \begin{equation}
   n^\mu \partial_\mu = 
    \frac{1}{\sqrt{1+\phi'^2}} (\phi' \partial_\rho - \partial_y) ,\quad
    n_\mu dx^\mu = \frac{1}{\sqrt{1+\phi'^2}}
    (\phi'd\rho - dy) .
  \end{equation}
  Also, those tangent to the brane are 
  \begin{equation}
   r^\mu \partial_\mu = 
    \frac{1}{\sqrt{1+\phi'^2}} (\partial_\rho + \phi' \partial_y) ,\quad
    r_\mu dx^\mu = \frac{1}{\sqrt{1+\phi'^2}}
    (d\rho + \phi' dy) .
  \end{equation}

  Non-zero components of the extrinsic curvature are 
  \begin{equation}
   K_{\mu\nu} r^\mu r^\nu =
    \left(\frac{\phi'}{\sqrt{1+\phi'^2}}\right)' ,\quad
    K^m{}_n = \frac{\phi'}{\rho\sqrt{1+\phi'^2}} \delta^m{}_n ,
  \end{equation}  
  where $m, n$ are running on the ($p-2$)-sphere.
  The stress-energy tensor of the brane is 
  \begin{equation}
   T^0{}_0 = T^1{}_1 = - \frac{1}{\sqrt{1-c^2\phi^2}}, \quad
    T_{\mu\nu}r^\mu r^\nu = - \sqrt{1-c^2\phi^2}, \quad
    T^m{}_n = - \sqrt{1-c^2\phi^2} \delta^m{}_n .
  \end{equation}
  It turns out that the tension on the cone, which is given by 
  $\sigma \equiv \sqrt{1-c^2\phi^2}$, becomes isotropic.

  Let us write the force balance explicitly.
  In this case there is an external force because of the NSNS field.
  The external force $\mathcal{F}^\mu$ is given by 
  \begin{equation}
   \mathcal{F}^\mu = \frac{1}{2}H^\mu{}_{ab}J^{ab} ,\quad
    J^{ab} \equiv - \sigma \gamma^{ac} h^{bd} B_{cd} .
  \end{equation}
  where $J^{ab}$ denotes the current coupled with the NSNS field strength.
  The force balance along the normal direction~(\ref{eq:extrinsic_eom})
  is written as 
  \begin{equation}
   \begin{aligned}
    T^{\mu\nu}r_\mu r_\nu K_{\alpha\beta}r^\alpha r^\beta
    + T^m{}_n K^n{}_m
    - \frac{1}{2}\sigma n^\mu H_{\mu\alpha\beta}
    \partial_a X^\alpha \partial_b X^\beta \gamma^{ac} h^{bd} B_{cd} = 0 ,
   \end{aligned}
  \end{equation}
This can be rewritten as 
  \begin{equation}
   - \sqrt{1-c^2\phi^2} 
    \left(\frac{\phi'}{\sqrt{1+\phi'^2}}\right)' 
    - (p-2) \sqrt{1-c^2\phi^2} \frac{\phi'}{\rho\sqrt{1+\phi'^2}}
    - \frac{c^2\phi}{\sqrt{1+\phi'^2}} \frac{1}{\sqrt{1-c^2\phi^2}} = 0.
    \label{eq:EOM_flat}
  \end{equation}
It turns out that this equation is nothing but the equation of motion of the D-brane.

  Assuming that near $\rho \sim 0$ 
  the brane becomes conical as 
  $\phi(\rho) - c^{-1} \sim \rho$ and the tension $\sigma$ vanishes, 
  the equation of motion can be reduced to 
  \begin{equation}
   (p-2) \sqrt{1-c^2\phi^2} \frac{\phi'}{\rho\sqrt{1+\phi'^2}} \simeq
    - \frac{c^2\phi}{\sqrt{1+\phi'^2}} \frac{1}{\sqrt{1-c^2\phi^2}} .
  \end{equation}
  As a result, we have 
  \begin{equation}
   2(p-2) \phi'^2|_{\rho=0} = 1 ,
  \end{equation}
 which is the solution we found before, (\ref{branecone}).

\subsection{Probe brane cone in AdS/CFT}
\label{sec:exAdS}

\subsubsection{Conic solution}

We shall see that the universality of the cone angle $\theta_{\rm cone}$
is quite broadly found, 
and here 
we demonstrate a calculation in a popular setup in 
AdS/CFT. It has been known~\cite{Mateos:2006nu,Frolov:2006tc} that there exists a ``critical embedding'' at which
a flavor D-brane in AdS/CFT correspondence exhibits a conical shape,
which serves as a phase boundary.
In the following, we will find that the cone angle takes the universal form 
(\ref{univangle}) for a conic D7-brane with electromagnetic field in the background
of AdS$_5$-Schwarzschild$\times S^5$ geometry.

The background metric with a generic temperature $T$ is given by
\begin{eqnarray}
ds^2 = \frac{1}{R^2} \frac{w^4+r_{\rm H}^4/4}{w^2}
\left[-h(w) dt^2 + dx_i^2\right]
+ \frac{R^2}{w^2} \left(dw^2 + w^2 d\Omega_5^2\right)\, ,
\label{metricAdS}
\end{eqnarray}
where $i=1,2,3$,  and we defined  
\begin{eqnarray}
h(w) \equiv \left(\frac{w^4-r_{\rm H}^4/4}{w^4+r_{\rm H}^4/4}\right)^2\, .
\end{eqnarray}
The temperature is related to usual horizon radius $r_{\rm H}$ as
\begin{eqnarray}
r_{\rm H} = \pi T R^2 = \pi T \sqrt{2\lambda} \alpha' \, .
\end{eqnarray}
At $T=0$, the geometry reduces to that of AdS$_5\times S^5$.

We consider a D7-brane probe action in this geometry, with a generic constant
electromagnetic field strength on the brane.
The action in the geometry is
\begin{eqnarray}
S &=& -{\cal T}_p \int \! d^{p+1}x
\sqrt{-\det(g_{ab} + 2\pi \alpha' F_{ab}  + \partial_a \phi^k \partial_b \phi^l g_{kl} )}  
\end{eqnarray}
where $a,b=0,1,\cdots,7$ are for the D7-brane worldvolume coordinates, and
$k,l=8,9$ are transverse coordinates. 
We define the transverse coordinates as
\begin{eqnarray}
w^2 = \rho^2 + L(\rho)^2
\end{eqnarray}
where $\rho^2 = (x^4)^2 + \cdots + (x^7)^2$. The function $L(\rho)$ describes 
the shape of the D7-brane in the geometry, and we assume a spherical symmetry
with respect to the radial coordinate $\rho$ on the D7-brane, as for the shape.
That is, we decompose
\begin{equation}
    dw^2 + w^2 d\Omega^2_5
     = d\rho^2 + dL^2 + \rho^2 d\Omega^2_3 + L^2 d\psi^2 ,
\end{equation}
   where $w^2 = \rho^2 + L^2$, and the embedding function is given by $L = L(\rho)$ and 
   $\psi = \text{const.}$.

Some calculations of the Lagrangian leads to the following expression for the
action,
\begin{eqnarray}
S &=& \mbox{const.} \int \! d^4 x \; \rho^3
\sqrt{\xi \; (1+(L')^2)}
\label{DBIxi}
\end{eqnarray}
where 
\begin{eqnarray}
\xi &\equiv& 
\left(\frac{w^4+r_{\rm H}^4/4}{w^4}\right)^4 h(w)
\nonumber \\
& &- \left(\frac{2\pi \alpha' R^2 (w^4+r_{\rm H}^4/4)}{w^6}\right)^2 \left(
\textbf{E}^2 - h(w) \textbf{B}^2\right)
-\left(\frac{2\pi \alpha' R^2}{w^2}\right)^4 (\textbf{E}\cdot\textbf{B})^2 \, .
\end{eqnarray}
By increasing $E_i$ in this expression, there exists a critical electric field
at which $\xi=0$. In other words, for fixed $B_i$, $E_i$ and the background
(temperature $T$), there exists $L$ at which $\xi=0$ is realized. Let us denote
that value of $L$ as $L_0$.
Then, expanding the scalar function $L(\rho)$ around $L=L_0$ as
\begin{eqnarray}
L(\rho) = L_0 + f(\rho), 
\end{eqnarray}
we can have a leading order action
\begin{eqnarray}
S &=& \mbox{const.} \int \! d^4 x \; \rho^3
\sqrt{f \; (1+(f')^2)}
\end{eqnarray}
which is exactly of the form (\ref{f1ff}) with $p=5$. 
Therefore, we again obtain a conic D-brane whose tip is at $L=L_0$,
\begin{eqnarray}
L(\rho) = L_0 + \frac{1}{\sqrt{6}}\rho + \mbox{higher.}\, 
\label{coneAdSsol}
\end{eqnarray}

\subsubsection{Force balance of the cone}

Let us study the force balance.
The induced metric on the D$7$-brane becomes 
   \begin{equation}
    h_{ab}dy^a dy^b
     = \frac{1}{R^2}\frac{w^4 + r^4_\mathrm{H}/4}{w^2}
     [- h(w)dt^2 + dx_i^2] + 
     \frac{R^2}{w^2} [(1+L'^2)d\rho^2 + \rho^2 d\Omega^2_3] \, .
   \end{equation}   
   The unit vector and $1$-form normal to the brane are given by 
   \begin{equation}
    n^\mu \partial_\mu = \frac{w}{R}\frac{1}{\sqrt{1+L'^2}}
     [L'\partial_\rho - \partial_L] ,\quad
     n_\mu dx^\mu = \frac{R}{w} \frac{1}{\sqrt{1+L'^2}}[L'd\rho - dL] ,
   \end{equation}
   and the unit vector and $1$-form tangent to the brane are 
   \begin{equation}
    r^\mu \partial_\mu = \frac{w}{R}\frac{1}{\sqrt{1+L'^2}}
     [\partial_\rho + L' \partial_L]
     ,\quad
     r_\mu dx^\mu = \frac{R}{w} \frac{1}{\sqrt{1+L'^2}}[d\rho + L' dL] .
   \end{equation}

   The extrinsic curvature for the direction with the normal vector $n^\mu$ is 
   \begin{equation}
    \begin{aligned}
     K^0{}_0 =& 
     \frac{1}{Rw}\left(\sqrt{h} + \frac{w}{2h}\frac{dh}{dw}\right)
     \frac{\rho L' - L}{\sqrt{1+L'^2}}
     ,\\
     K^i{}_j =& 
     \frac{\sqrt{h(w)}}{Rw} \frac{\rho L' - L}{\sqrt{1+L'^2}} \delta^i{}_j
     , \\
     K^m{}_n =& \frac{1}{Rw}\frac{\rho + LL'}{\sqrt{1+L'^2}}\frac{L}{\rho}
     \delta^m{}_n ,
    \end{aligned}
   \end{equation}
   and 
   \begin{equation}
    K_{\mu\nu}r^\mu r^\nu = 
     \frac{w^2}{R}\left(\frac{L'}{w\sqrt{1+L'^2}}\right)'
     + \frac{L\sqrt{1+L'^2}}{Rw} .
   \end{equation}

   Non-vanishing components of the stress-energy tensor of the D$7$-brane are 
   \begin{equation}
    \begin{aligned}
     T^0{}_0 =& - \frac{1}{\sigma} (1+ \frac{\textbf{B}^2}{g^2}) 
     ,\quad 
     T^0{}_i = \frac{1}{\sigma} \frac{1}{g^2 h}
     \epsilon_{ijk}E^j B^k, \quad 
     T^i{}_0 = - \frac{1}{\sigma} \frac{1}{g^2}
     \epsilon^{ijk}E_j B_k
     ,\\
     T^i{}_j =& - \frac{1}{\sigma}
     \left[
     (1-\frac{\textbf{E}^2}{g^2 h})\delta^i{}_j
     + \frac{1}{g^2} (h^{-1}E^iE_j + B^i B_j)
     \right] ,\\
     T^m{}_n =& - \sigma \delta^m{}_n, \quad
     T_{\mu\nu}r^\mu r^\nu = - \sigma ,
    \end{aligned}
   \end{equation}
   where 
   \begin{equation}
    \sigma^2 \equiv 1 - \frac{1}{g^2}(h^{-1}\textbf{E}^2 - \textbf{B}^2) 
     - \frac{1}{g^4 h} (\textbf{E}\cdot\textbf{B})^2 ,
   \end{equation}
   and 
   \begin{equation}
    g(w) \equiv 
     \frac{1}{2\pi \alpha' R^2} \frac{w^4 + r_\mathrm{H}^4/4}{w^2} .
   \end{equation}
   They imply that, if the electric field becomes sufficiently large to
   be $\sigma = 0$, some components of the stress-energy tensor, which mean the
   isotropic tension,  will vanish and
   the others will diverge. 

   From Eq.~(\ref{eq:extrinsic_eom}), the equation of motion for the brane can be written as 
   \begin{equation}
    T^0{}_0 K^0{}_0 + T^i{}_j K^j{}_i + T^m{}_n K^n{}_m
     + T^{\mu\nu}K_{\alpha\beta} r_\mu r_\nu r^\alpha r^\beta = 0 .
   \end{equation}
   Now, we suppose that the brane shape will become conical at the critical
   point, namely $\sigma = 0$ at $\rho = 0$.
   Since $T_{\mu\nu}r^\mu r^\nu$ vanishes at $\sigma=0$, 
   the equation of motion reduces to 
   \begin{equation}
    T^0{}_0 K^0{}_0 + T^i{}_j K^j{}_i \simeq - T^m{}_n K^n{}_m ,
     \label{eq:balance_D7}
   \end{equation}
   which means force balance at the tip of the cone.
   It turns out that the stress-energy tensor in the left-hand side will diverge at
   $\sigma = 0$ while the extrinsic curvature in the right-hand side
   will diverge because of $\rho = 0$.
   In contrast to the previous two examples, there is no explicit
   external force now.
   However, gravitational force due to the bulk AdS space is acting on
   the brane and balanced with the tension coupling the extrinsic
   curvature of the cone.
   By using $\sigma(\rho=0) = 0$ Eq.~(\ref{eq:balance_D7}) can be
   explicitly rewritten as 
   \begin{equation}
    - \frac{1}{R}
     \frac{d\sigma}{dw} \frac{\rho L' - L}{\sqrt{1+L'^2}}
    \simeq
    3 \sigma\frac{1}{Rw}\frac{\rho + LL'}{\sqrt{1+L'^2}}\frac{L}{\rho}.
   \end{equation}
   
   As a result, we have 
   \begin{equation}
    L'|_{\rho=0} = \frac{1}{\sqrt{6}} ,
   \end{equation}
   where we have used 
   $\sigma / \rho|_{\rho=0} = 2w'd\sigma/dw|_{\rho=0}$ and 
   $w|_{\rho=0} = L_0$.
   This is equivalent to the conic solution (\ref{coneAdSsol}).


\subsection{Universal cone angle}
\label{cone-angle}

The Taylor cones have a universal cone angle. We have seen that the 
mechanism of the formation of the conic D-branes is quite similar to that of the 
Taylor cones, thus we expect that there may exist a universal cone angle
for the D-brane cones.

In fact, we find that the half-cone angle is universally determined as
\begin{eqnarray}
\theta_{\rm cone} = \arctan \sqrt{2(d_{\rm cone}-1)} \, ,
\label{univangle}
\end{eqnarray}
where $d_{\rm cone}$ is the dimension of the cone (in other words, the
cone is ${\bf R}_+ \times S^{d_{\rm cone}-1}$). To show this, we first look
at the example in the NSNS background in Sec.~\ref{sec:exNSNS}. 
There 
the D$p$-brane has the
cone in the worldvolume directions $(x^2, x^3, \cdots, x^p)$, so
the cone is $(p-1)$-dimensional : $d_{\rm cone}= p-1$. From
the cone solution (\ref{branecone}), the half-cone angle defined
as
\begin{eqnarray}
\tan \theta_{\rm cone} = \left(\frac{d\phi}{d\rho}\right)^{-1}
\biggm|_{\rho=0}
\end{eqnarray} 
is given by (\ref{univangle}). 

This angle (\ref{univangle}) of Sec.~\ref{sec:exNSNS} 
should be quite universal, since 
the linear part of the solution (\ref{branecone})
does not depend on the value $c$ of the background NSNS flux. 
Let us check the universality of the cone angle below.

For the D-brane cone in the RR background in Sec.~\ref{sec:exRR},
the dimension of the cone is obviously $d_{\rm cone}=2$, so the formula
(\ref{univangle}) suggests $\theta_{\rm cone}=\arctan\sqrt{2}$. This coincides
with the solution (\ref{uniqueab}) in the RR background.
Again, the angle does not depend on the strength of the RR flux, so the
angle is universal.

Furthermore, as for the D7-brane cone in AdS in Sec.~\ref{sec:exAdS}, 
the cone angle is found to be
\begin{eqnarray}
\theta_{\rm cone} = \arctan \sqrt{6}
\label{coneAdS}
\end{eqnarray}
The probe D7-brane
has 4 spatial dimensions for its worldvolume in the extra dimensions,
since the worldvolume along $(x^0,\cdots,x^3)$ is 
assumed to be flat as a quark flavor D-brane.
So the current situation  corresponds to the case of $d_{\rm cone}=4$. 
Therefore this (\ref{coneAdS}) 
coincides again with the universal cone angle formula (\ref{univangle}).
Again, cone angle of the D7-brane in AdS 
is independent of all the background values: the 
black hole temperature $T$, the magnetic field $B_i$, the critical 
electric field $E_i$, and the position of 
the cone tip $L_0$.

Note that in the last case, 
the angle, of course, should be given by the inner product associated
with the bulk metric between tangent vectors on the brane worldvolume,%
\footnote{
The angle between vectors $u$ and $v$ on the metric $g$ is defined by 
$\theta(u,v) = \arccos 
\frac{g(u,v)}{\sqrt{g(u,u)}\sqrt{g(v,v)}}$ where 
$g(u,v) \equiv u^\mu v^\nu g_{\mu\nu}$. 
The angle and the length are geometrically independent quantities,
because, for example, under a conformal transformation $g \to \Omega^2 g$ the angle is invariant but the
length is not.} 
so that the value does not depend on the choice of the coordinates.
Now, the metric (\ref{metricAdS}) which we use is conformal to Euclidean
space in a Cartesian coordinate system for the extra dimensions,
as seen from the factor $(dw^2 + w^2 d\Omega_5^2)$.
The expression of the angle is the same as that in the flat space.

From all the examples we studied above,  
the half-cone angle (\ref{univangle}) is independent of the 
background parameters. 
In comparison with the general formula (\ref{eq:general_formula}) discussed in Sec.~\ref{sec:covariant}, a factor of $(d_\mathrm{cone}-1)$ obviously comes from the
dimension of the spherical part of the cone, and the other factor of two
is related to the power of stress distribution on the cone, that is, the
tension on the cone behaves $r^\alpha$ with $\alpha = 1/2$ at the
distance $r$ from the apex. 
We conjecture that the D-brane cone angle (\ref{univangle})
is independent of anything related to the background fields. 
The universality is reminiscent of the Taylor cones.

  

\section{Universality of spectra in AdS/CFT}

\subsection{Observables in boundary theory}

In section~\ref{sec:exAdS}, we have seen that the cone angle is
universal for the probe D7-brane in the AdS$_5$-Schwarzschild$\times S^5$ 
spacetime: 
The cone angle does not depend on the parameters, 
Hawking temperature $T$, electric field $E$ and magnetic field $B$, 
once we impose the critical condition on the tip of the brane.
The D3/D7 system is dual to $\mathcal{N}=2$ super symmetric QCD.
Here, we will investigate 
how we can observe the universality of the cone angle in the view of dual gauge theory.
A static D7-brane solution in the bulk spacetime
is written as $L=L(\rho)$.
Near the AdS boundary $\rho=\infty$, the solution is expanded as
\begin{equation}
 L(\rho)=L_\infty + \frac{c}{\rho^2}+\cdots\ .
\end{equation}
The constants $L_\infty$ and $c$ correspond to quark mass $m_q$ and
quark condensate $\langle \mathcal{O} \rangle$ as
\begin{align}
&m_q=\left(\frac{\lambda}{2\pi^2}\right)^{1/2}\frac{L_\infty}{R^2}\ ,\label{mq}\\
&\langle \mathcal{O} \rangle
=- \frac{N_c \sqrt{\lambda}}{2^{3/2} \pi^3} \frac{c}{R^6}\ .\label{quarkcond}
\end{align}
The other observable in the boundary theory is 
the energy density.
The quark condensate and energy density are summations of 
contributions from all meson excitations.
To obtain expressions for each meson excitation,
we will study the linear perturbation theory in the following subsections.
We will find that 
the universality of the cone angle is observed as 
the universality of the spectra of quark condensate and energy density.

We are also motivated by the ``turbulent meson condensation'' proposed by
Refs.\cite{Hashimoto:2014xta,Hashimoto:2014dda}. 
We have numerically studied the time evolution of energy spectrum in  dynamical and quasi-static
processes in the D3/D7 system and found that, as far as we investigated,
the energy spectrum always obeys power law, $\varepsilon_n\sim n^{-5}$,
at the phase boundary of the
black hole and Minkowski embeddings.
So, we conjectured that the exponent $-5$ in the energy spectrum 
is universal for phase transitions in the $\mathcal{N}=2$ SQCD.
We will give an analytic derivation of the exponent 
for a quasi-static process in this section.

\subsection{Linear perturbation theory}

For the zero temperature $r_H=0$, the background
metric~(\ref{metricAdS}) reduces to the AdS$_5\times S^5$ spacetime:
\begin{eqnarray}
ds^2 = \frac{w^2}{R^2}
(-dt^2 + dx_i^2)
+ \frac{R^2}{w^2} (d\rho^2 +\rho^2d\Omega_3^2
+dL^2 + L^2 d\psi^2)\, ,
\label{pureAdS}
\end{eqnarray}
where $w^2=\rho^2+L^2$. We impose the spherical symmetry of $S^3$ and
translation invariance $\partial_{x_i}$ on the D7-brane. Then, 
dynamics of the D7-brane in this spacetime is described by a single function
$L=L(t,\rho)$. 
The static solution is trivially given by a constant: $L(t,\rho)=L_\infty$.
We consider linear perturbation of the static solution.
Defining the perturbation variable $\delta L=L(t,\rho)-L_\infty$,
we obtain second order action for $\delta L$ as
\begin{equation}
 S=\pi^2 \mathcal{T}_7 R^4 V_3 \int dt d\rho \frac{\rho^3}{(\rho^2+L_\infty^2)^2}
\left[
\dot{\delta L}^2-\frac{(\rho^2+L_\infty^2)^2}{R^4}\delta L'{}^2
\right]\ ,
\label{secaction}
\end{equation}
where ${}^\cdot=\partial_t$, ${}'=\partial_\rho$, 
$V_3= \int dx_1 dx_2 dx_3$ and 
$\mathcal{T}_7=(2\pi)^{-7}\alpha'{}^{-4}g_s$.
The equation of motion for $\delta L$ is 
\begin{equation}
 (\partial_t^2 + \mathcal{H})\delta L=0\ ,\quad
 \mathcal{H}=-\frac{(\rho^2+L_\infty^2)^2}{R^4\rho^3}\partial_\rho
 \rho^3\partial_\rho\ .
\label{EQ}
\end{equation}
The operator $\mathcal{H}$ is Hermitian under an inner product,
\begin{equation}
 (\alpha,\beta)=\int_0^\infty d\rho \frac{\rho^3}{(\rho^2+L_\infty^2)^2} \alpha\beta\ .
\label{innprod}
\end{equation}
We define the norm using the inner product as $\|\alpha\|^2=(\alpha,\alpha)$.
Eigenvalues $\omega_n^2$ and eigen functions $e_n$ of $\mathcal{H}$ are
given by
\begin{equation}
e_n=\mathcal{N}_n\frac{L_\infty^2}{\rho^2+L_\infty^2}
 F\left(n+3,-n,2;\frac{L_\infty^2}{\rho^2+L_\infty^2}\right)\ ,\quad
\omega_n^2=4(n+1)(n+2)\frac{L_\infty^2}{R^4}\ .
\end{equation}
where $n=0,1,2,\cdots$ and 
$\mathcal{N}_n=\sqrt{2(2n+3)(n+1)(n+2)}$.
The eigen functions are normalized as
$(e_n,e_m)=\delta_{nm}$.
We expand the perturbation variable as
\begin{equation}
 \delta L=\sum_{n=0}^\infty c_n(t)e_n(\rho)\ .
\label{modedec}
\end{equation}
The asymptotic form of the eigen function is 
$e_n\simeq \mathcal{N}_n L_\infty^2/\rho^2$ ($\rho\to\infty$). Thus, from
Eq.~(\ref{quarkcond}), the quark condensate for the fluctuating D7-brane 
is written as
\begin{equation}
 \langle \mathcal{O}\rangle=\sum_{n=0}^\infty \langle
 \mathcal{O}_n\rangle\ ,\qquad
 \langle \mathcal{O}_n\rangle
=- \frac{\mathcal{N}_n N_c m_q^3}{\lambda} \frac{c_n(t)}{L_\infty}\ ,
\end{equation}
where $\langle \mathcal{O}_n\rangle$
can be regarded as the quark condensate contributed from $n$-th
excited mesons. 
We eliminated the AdS curvature scale $R$ using Eq.~(\ref{mq}). So, the quark mass $m_q$
appeared in the expression of $\langle \mathcal{O}_n\rangle$. From the second order action~(\ref{secaction}), we obtain the
energy density as 
\begin{equation}
 \varepsilon=
\pi^2 \mathcal{T}_7 R^4 \int d\rho \frac{\rho^3}{(\rho^2+L_\infty^2)^2}
\left[
\dot{\delta L}^2+\frac{(\rho^2+L_\infty^2)^2}{R^4}\delta L'{}^2
\right]\ ,
\label{engy}
\end{equation}
Substituting Eq.~(\ref{modedec}) into above expression, we have
\begin{equation}
 \varepsilon=\sum_{n=0}^\infty \varepsilon_n\ ,\qquad
\varepsilon_n=\frac{N_c m_q^2}{8\pi^2 L_\infty^2}(\dot{c}_n^2+\omega_n^2 c_n^2)\ ,
\end{equation}
We substituted $\alpha'{}^2=R^4/(2\lambda)$ and eliminated $R$ using Eq.~(\ref{mq}) again. $\varepsilon_n$ can be
regarded as the energy contribution from $n$-th excited meson.

\subsection{Spectra in non-linear theory}

We extend definitions of spectra $\langle \mathcal{O}_n\rangle$ and
$\varepsilon_n$ to non-linear theory.
Denoting a static non-linear D7-brane solution in the
AdS$_5$-Schwarzschild$\times S^5$ spacetime
as $L=L(\rho)$, 
we define spectra of quark condensate and energy density for the
non-linear solution as
\begin{equation}
  \langle \mathcal{O}_n\rangle
=- \frac{\mathcal{N}_n N_c m_q^3}{\lambda L_\infty} (L-L_\infty,e_n)\
,\qquad
\varepsilon_n=\frac{N_c m_q^2 \omega_n^2}{8\pi^2L_\infty^2} (L-L_\infty,e_n)^2\ ,
\label{cond_nonlin}
\end{equation}
where we have omitted time derivative of the mode coefficient
$\dot{c}_n$ in Eq.~(\ref{engy}) since we consider only static solutions in
non-linear theory.
We will see that the universality of the cone angle can be seen as the
universality of spectra in the large-$n$ limit.

For the purpose of studying the large-$n$ behavior of the spectra, 
it is convenient to show the following lemma:\\
\textbf{Lemma} 
\textit{
Let $f(\rho)$ be a smooth function in $\rho\in (0,\infty)$
whose norm is finite, i.e, $\|f\|^2<\infty$.\footnote{
The condition for the finiteness of the norm can be explicitly written as 
$f=o(1/\rho^2)$ ($\rho\to 0$) and
$f=o(1)$ ($\rho\to \infty$),
where $o$ is the Landau's Little-o notation.
} 
Then,  }
\begin{equation}
 (f,e_n)\to 0\ ,\qquad (n\to \infty)\ .
\label{RL}
\end{equation}
\textbf{Proof: } 
{}From the positivity of the norm, we obtain
\begin{multline}
0\leq \| f-\sum_{n=0}^N (f,e_n)e_n \|^2\\
=
\|f\|^2-2\sum_{n}(f,e_n)^2+\sum_{n,m}(f,e_n)(f,e_m)(e_n,e_m)
=\|f\|^2-\sum_{n}(f,e_n)^2\ .
\end{multline}
At the last equality, we used the orthonormal relation $(e_n,e_m)=\delta_{nm}$.
Hence, we have
\begin{equation}
 \sum_{n=0}^N(f,e_n)^2 \leq \|f\|^2\ .
\end{equation}
This inequality is known as Bessel's inequality.
Now, the right hand side of the inequality is finite and does
not depend on $N$. Therefore, $(f,e_n)$ must decrease to zero for 
$n\to \infty$. $\Box$

In the Fourier analysis, this is well known as Riemann-Lebesgue lemma.

\subsection{Spectra of non-conical solutions}

Firstly, we investigate the large-$n$ behavior of the spectra for 
non-conical solutions.
Solving the equation of motion derived from the DBI action~(\ref{DBIxi}) 
near the axis $\rho=0$, the non-conical solution is expanded as
\begin{equation}
 L(\rho)=a_0 + a_2\rho^2 + a_4 \rho^4 + \cdots\ ,\quad
  (\rho\sim 0)
\end{equation}
On the other hand, at the infinity,
the solution behaves as
\begin{equation}
 L(\rho)-L_\infty=\frac{b_{2}}{\rho^2} + \frac{b_{4}}{\rho^4}+
  \frac{b_{6}}{\rho^6}+\cdots\ ,\quad (\rho\sim \infty)\ .
\label{reg_inf}
\end{equation}
Operating $\mathcal{H}$ to the non-linear solution $m$-times, we
have
\begin{equation}
 \mathcal{H}^mL =\mathcal{O}(1)\ ,\quad (\rho\sim 0)\ ,\qquad
 \mathcal{H}^mL =\mathcal{O}(1/\rho^2) ,\quad (\rho\sim \infty)\ .
\end{equation}
So, $\|\mathcal{H}^m L\|^2$ is finite for any $m$.
Therefore, from Eq.~(\ref{RL}), 
we obtain 
\begin{equation}
 (\mathcal{H}^m L,e_n)
=(L-L_\infty,\mathcal{H}^me_n)
=\omega_n^{2m}(L-L_\infty,e_n)
\to 0 \ ,\quad (n\to \infty)\ .
\end{equation}
At the first equality, we used hermiticity of $\mathcal{H}$.
This implies that $(L-L_\infty,e_n)$ must fall off faster than 
$1/\omega_n^{2m}\sim n^{-2m}$ as a function of $n$.
Since $m$ is any integer, $(L-L_\infty,e_n)$ must fall off faster than any
power law functions of $n$. Therefore, spectra of quark condensate and energy
density also fall off faster than any power.
Actually, by the numerical calculation, it
is suggested that 
spectra fall off exponentially for non-conical solutions~\cite{Hashimoto:2014xta,Hashimoto:2014dda}.

\subsection{Spectra of conical solutions}

Now, we consider the mode decomposition of a conical solution $L=L_c(\rho)$. 
Near the axis $\rho=0$, the critical solution is expanded as
\begin{equation}
 L_c(\rho)=a_0 +\cot\theta_\textrm{cone}\,\rho + a_2\rho^2 + a_3\rho^3 + \cdots\
  ,\quad(\rho\sim 0)\ ,
\end{equation}
where $\theta_\textrm{cone}$ represents the cone angle.
Asymptotic form at the infinity is same as that for the non-conical 
solution~(\ref{reg_inf}). 
We define $\tilde{L}_c$ which satisfies Neumann condition at $\rho=0$ as
\begin{equation}
 L_c(\rho)=\tilde{L}_c(\rho)+L_\infty S(\rho)\cot\theta_\textrm{cone}\ .
\label{rs_dec}
\end{equation}
where we define the function $S$ as
\begin{equation}
 S(\rho)\equiv \frac{L_\infty^2\rho}{(\rho^2+L_\infty^2)^{3/2}}\ .
\end{equation}
We chose this function so that we can carry out its mode
decomposition analytically as\footnote{
We expressed the eigen function $e_n$ by the series of $L_\infty^2/(\rho^2+L_\infty^2)$
and integrated termwise.
}
\begin{equation}
\begin{split}
 (S,e_n)
&=\frac{6(-1)^{n+1}\sqrt{2(2n+3)(n+1)(n+2)}}
{(2n-1)(2n+1)(2n+3)(2n+5)(2n+7)}\\
&\simeq \frac{3}{8}(-1)^{n+1}n^{-7/2}\ , \qquad\qquad\qquad(n\to \infty)\ .
\end{split}
\end{equation}

Next we consider the mode decomposition of $\tilde{L}_c(\rho)$.
The asymptotic form of $\tilde{L}_c$ becomes
\begin{equation}
 \tilde{L}_c(\rho)=a_0 + a_2\rho^2 + \left(a_3+\frac{3}{2L^2}\cot\theta_\textrm{cone}\right)\rho^3 + \cdots\ ,\quad
  (\rho\sim 0)\ .
\label{tilchi_as}
\end{equation}
There is no linear term in $\rho$ since it is subtracted
by $L_\infty S(\rho) \cot \theta_\textrm{cone}$ in Eq.~(\ref{rs_dec}). 
Operating $\mathcal{H}$ twice to $\tilde{L}_c(\rho)$, 
we obtain 
\begin{equation}
 \mathcal{H}^2\tilde{L}_c=\mathcal{O}(1/\rho)\ ,\quad (\rho\sim 0)\ ,\qquad
 \mathcal{H}^2\tilde{L}_c =\mathcal{O}(1/\rho^2) ,\quad (\rho\sim \infty)\ .
\end{equation}
Although $\mathcal{H}^2\tilde{L}_c$ diverges at $\rho=0$, its norm 
is still finite. (Because of
the measure in the inner product~(\ref{innprod}), the integrand is
regular at $\rho=0$.)
From Eq.~(\ref{RL}), we obtain
\begin{equation}
  (\mathcal{H}^2\tilde{L}_c,e_n)=\omega_n^4(\tilde{L}_c - L_\infty,e_n)\to 0 \ ,\quad (n\to \infty)\ .
\end{equation}
So, $(\tilde{L}_c - L_\infty,e_n)$ falls off faster than
$1/n^4$ and is a subleading term in the limit of $n\to \infty$.
Thus, we have
\begin{equation}
 (L_c-L_\infty,e_n)\simeq \frac{3}{8}(-1)^{n+1}n^{-7/2} L_\infty\cot\theta_\textrm{cone} \ ,\quad (n\to
  \infty)\ .
\end{equation}
Therefore, from Eq.~(\ref{cond_nonlin}), spectra of quark condensate and
energy density become
\begin{equation}
  \langle \mathcal{O}_n\rangle
\simeq \frac{3N_c m_q^3 \cot\theta_\textrm{cone}}{4\lambda} 
\frac{(-1)^n}{n^{2}}\ ,\quad 
\varepsilon_n
\simeq \frac{9N_c m_q^4 \cot^2\theta_\textrm{cone}}{64\lambda} 
\frac{1}{n^5}\ ,\quad 
\quad (n\to \infty)\ .
\end{equation}
There are some remarks on the large-$n$ limit of spectra.
The spectrum of the quark condensate is positive/negative when $n$ is an
even/odd number. This is one of the simplest prediction from the AdS/CFT
calculation of the spectrum.
The power of spectra does not depend on the cone angle
$\theta_\textrm{cone}$. 
So, the universality of the cone angle has nothing to do with
the universality of the power in the spectra: 
If the D7-brane has a cone, 
we always have 
$\langle \mathcal{O}_n\rangle\propto n^{-2}$ and 
$\varepsilon_n \propto n^{-5}$ for $n\to \infty$. 
The universality of the cone angle however 
appears as the universality of the coefficient of the power law behavior in the spectra.
This could be one of the observable effects in the dual gauge theory.
It would be nice if these predictions from AdS/CFT can be confirmed by
lattice QCD.

\section{Conclusion and discussion}

In this paper, we found conic D-brane solutions under external uniform 
RR or NSNS field strengths. The apex angle is found to be universal. 
The angle formula (\ref{coneangle1}) depends only on the dimensions
of the cone.

As we emphasized, the universal angle of the D-brane cone 
is similar to the one for Taylor cones in fluid mechanics. The D-brane 
mechanics is governed by the DBI action which normally exhibits
the form such as (\ref{effectiveL}) when the tension goes to zero at a certain point
on the worldvolume. This special form (\ref{effectiveL}) is important
for having the conic shape and the universal angle.

When D-branes touch an event horizon in the target space, a cone
forms and it is nothing but the critical embedding at thermal phase transition in AdS/CFT.
However, the situation is different from
our cases. 
For example, the special form (\ref{effectiveL}) consists of an
overall factor $\sqrt{f}$ related to the tension and the Nambu-Goto part 
$\sqrt{1 + (f')^2}$ coming from simply the volume element of the brane.
For a thermal phase transition \cite{Mateos:2006nu,Frolov:2006tc}, 
the conic shape appears due to the
nontrivial background spacetime and the deformation of the Nambu-Goto
part other than the tension.
For our cases the background geometry does not help anything, while the
coupling to the flux generates the vanishing factor $\sqrt{f}$ 
which results in cancellation of the tension.
In fact, in general for DBI actions the whole Lagrangian is written with a square root,
so when the factor goes to zero, it naturally vanishes as $\sqrt{f}$ for $f \rightarrow 0$.
This dependence determines the power 
of distribution of the tension as $\alpha = 1/2$ in (\ref{assumealpha})
and thus leads to the factor $2$ in the apex angle formula (\ref{coneangle1}).
This plausible argument is a support for genericity of the angle formula (\ref{coneangle1})
for any D-brane cones caused by external uniform forces.

The vanishing stress of the conic D-brane at its apex, is similar to D-brane
super tubes \cite{Mateos:2001qs} and tachyon condensation \cite{Sen:1998sm}.
In both cases, peculiar dispersions for propagation modes were reported
\cite{Palmer:2004gu,Gibbons:2002tv}. 
It would be interesting to study the modes around the apex of our conic D-branes.
One possible obstacle would be higher derivative corrections. In our examples,
since the apex angle is universal, it is impossible to gradually change the apex angle.
So the effects of the higher derivative terms can be non-infinitesimal. 

In a sense, our example of the D2-D0 cone in the RR 2-form flux
contrasts a renowned Myers effect \cite{Myers:1999ps}
which is for D2-D0 in a RR 4-form flux, the dielectric branes. The Myers effect
is a dielectric polarization of a D2-brane which is caused by the 4-form flux
pulling the D2-brane surface. In our case the 2-form flux pulls the bound D0-brane
on the surface of the D2-brane, so the mechanism of the forces is different, 
and resultantly, the shape of the D2-brane is different: a spherical or a conic shape.
It is obvious that these two effects can be combined once we allow both the 2-form and
the 4-form. More complicated brane shape can emerge, which may be important
for various applications of D-brane physics.

Finally, we would like to make a comment on spiky singularities in membrane quantization
\cite{deWit:1988ct}. We are not sure if the conic D-brane configurations may help resolving the issue or not. However one important observation is that, since the apex angle is unique,
the conic brane configuration can exist 
even if we turn off gradually the background flux.  This fact may signal a possible classification of classical D-brane configurations. We hope that this direction of research
may help for the quantization problem.

\acknowledgments
The work of K.~H.~was supported in part by JSPS KAKENHI Grant Numbers 15H03658, 15K13483.
This work of K. M. was supported by JSPS KAKENHI Grant Number 15K17658.

\appendix

\section{Stress-energy tensor of D-branes}

We consider a D$p$-brane in $d$-dimensional spacetime.
Let $\{x^\mu\}$ and $\{y^a\}$ be coordinates on the bulk spacetime and
the brane worldvolume, respectively.
The brane is characterized by embedding functions $x^\mu = X^\mu(y)$.
The DBI action for the D$p$-brane is given by 
\begin{equation}
 \begin{aligned}
  S_{\mathrm{D}p} =&
  - \int d^{p+1}y \sqrt{-\det(h_{ab} + F_{ab})} \\
  =& - \int d^{p+1}y \sqrt{-\det h_{ab}} \, \sigma ,
 \end{aligned}
\end{equation}
where 
\begin{equation}
 \sigma^2 \equiv \sum_{k=0}^{\lfloor (p+1)/2 \rfloor} 
  F_{a_1}{}^{b_1} \cdots F_{a_{2k}}{}^{b_{2k}}
  \delta_{[b_1}{}^{a_1}\cdots \delta_{b_{2k}]}{}^{a_{2k}} .
\end{equation}
Note that $h_{ab}$ denotes the induced metric, which is used for raising
and lowering Latin indices.
If the rank of $F_{ab}$ is less than or equal to three, we have
explicitly $\sigma^2 = 1 + \frac{1}{2}F_{ab}F^{ab}$.
If the rank of $F_{ab}$ is less than or equal to five, we have 
$\sigma^2 = 1 + \frac{1}{2}F_{ab}F^{ab}
 -
 \frac{1}{4}F_{a_1}{}^{a_2}F_{a_2}{}^{a_3}F_{a_3}{}^{a_4}F_{a_4}{}^{a_1}
+ \frac{1}{8}(F_{ab}F^{ab})^2$ .

The stress-energy tensor of the D$p$-brane in the bulk is given by 
\begin{equation}
 \begin{aligned}
  \hat{T}^{\mu\nu} =& 
  \frac{2}{\sqrt{-g}}\frac{\delta S_{\mathrm{D}p}}{\delta g_{\mu\nu}} \\
  =& -\frac{1}{\sqrt{-g}}\int d^{p+1}y \sqrt{-h} \, \sigma \, (\gamma^{-1})^{ab} 
  \partial_a X^\mu \partial_b X^\nu \delta^{(d)}(x-X(y)) ,
 \end{aligned}
\end{equation}
where $(\gamma^{-1})^{ab}$ denotes the inverse of the effective metric
defined by $\gamma_{ab} \equiv h_{ab} + F_{ac}F_{bd}h^{cd}$.
Here, we introduce $(d-p-1)$ functions $z^i(x)$ in the bulk such that 
$z^i(X(y)) = z_0^i$ satisfy, where arbitrary constants $z_0^i$ can be zero without loss of generality.
We have a diffeomorphism $x^\mu = \varphi^\mu(y^a, z^i)$ such that
$X^\mu(y^a) = \varphi^\mu(y^a,z^i=0)$.
This means that neighborhood of the brane is locally spanned by new bulk
coordinates $\{y^a,z^i\}$ and submanifold of the brane is
characterized by $z^i=0$.
The bulk metric can be rewritten as 
\begin{equation}
 g_{\mu\nu}dx^\mu dx^\nu = h_{ab} dy^a dy^b + N_{ij}dz^idz^j .
\end{equation}
As a result, we have 
\begin{equation}
 \begin{aligned}
  \hat{T}^{\mu\nu}
  =& -\frac{1}{\sqrt{-g(x)}}\int d^{p+1}y \sqrt{-h} \, \sigma \, (\gamma^{-1})^{ab} 
  \partial_a X^\mu \partial_b X^\nu \delta^{(d)}(x-X(y)) \\
=& -\frac{1}{\sqrt{N(y',z')}}\int d^{p+1}y \, \sigma \, (\gamma^{-1})^{ab} 
  \partial_a X^\mu \partial_b X^\nu 
  \delta^{(p+1)}(y'-y) \delta^{(d-p-1)} (z') \\
  =& -\frac{1}{\sqrt{N(y',z')}} \sigma \, (\gamma^{-1})^{ab} 
  \partial_a X^\mu \partial_b X^\nu \delta^{(d-p-1)} (z') ,
 \end{aligned}
\end{equation}
where we have used $\delta^{(d)}(x)/\sqrt{-g} = \delta^{(p+1)}(y)
\delta^{(d-p-1)}(z)/\sqrt{-hN}$ and $N\equiv \det N_{ij}$.

Now, the stress-energy tensor localized on the brane is obtained by
integrating $\hat{T}^{\mu\nu}$ along the directions perpendicular to the
brane as   
\begin{equation}
 \begin{aligned}
 T^{ab} =& \int d^{d-p-1}z \sqrt{N} \hat{T}^{\mu\nu}h^{a}{}_\mu
  h^b{}_\nu \\
  =& - \sigma (\gamma^{-1})^{ab} .
 \end{aligned}
\end{equation}
It is worth noting that this is equivalent to the stress-energy tensor
derived from variation of the induced metric on the worldvolume as 
\begin{equation}
 \begin{aligned}
  T^{ab} =& \frac{2}{\sqrt{-h}}
  \frac{\delta S_{\mathrm{D}p}}{\delta h_{ab}} \\
  =& - 2 \frac{\delta \sigma}{\delta h_{ab}}
  - \sigma h^{ab}
  = - \sigma (\gamma^{-1})^{ab} .
 \end{aligned}
 \label{eq:DEF_energy_momentum}
\end{equation}
This expression makes it clear that, since $\sigma$ is not
constant but depends on the induced metric for DBI branes, the energy
density is not equal to the tension (i.e. negative pressure) rather than
Nambu-Goto branes.


\end{document}